\documentclass[12pt]{article}
\usepackage{graphicx}
\usepackage{axodraw}
\usepackage[dvips]{color}
\usepackage{epsfig}
\newcommand{\slet}[1]{#1\!\!\! \slash}
\def\vmol{v_{\rm{M\slet{o}l}}}
\def\Journal#1#2#3#4{{#1} {\bf #2}, #3 (#4)}
\def\beq{\begin{equation}}
\def\eeq{\end{equation}}

\def\NPB{{\em Nucl. Phys.} B}
\def\PLB{{\em Phys. Lett.}  B}
\def\PRL{\em Phys. Rev. Lett.}
\def\PRD{{\em Phys. Rev.} D}

\def\r2{\sqrt 2}
\def\beq{\begin{equation}}
\def\eeq{\end{equation}}
\def\beqn{\begin{eqnarray}}
\def\eeqn{\end{eqnarray}}

\def\sinW2{\sin^2\theta_W}

\def\mz2{M_{z}^2}
\def\c2b{\cos 2\beta}

\def\m#1{{\tilde m}_#1}

\def\mz{M_z}

\def\Fq2{F_{2}(q^2)}
\def\f{\({\cal F}\)}
\def\d1{{\f(\tilde c;\tilde s;\tilde W)+ \f(\tilde c;\tilde \mu;\tilde W)}}

\def\sec2w{sec^2\theta_W}
\def\m12{m_{\frac{1}{2}}}
\def\12{$\frac{1}{2}$}
\def\m0{$m_0$}

\begin{document}
\baselineskip 18pt
\def\today{\ifcase\month\or
 January\or February\or March\or April\or May\or June\or
 July\or August\or September\or October\or November\or December\fi
 \space\number\day, \number\year}
\def\thebibliography#1{\section*{References\markboth
 {References}{References}}\list
 {[\arabic{enumi}]}{\settowidth\labelwidth{[#1]}
 \leftmargin\labelwidth
 \advance\leftmargin\labelsep
 \usecounter{enumi}}
 \def\newblock{\hskip .11em plus .33em minus .07em}
 \sloppy
 \sfcode`\.=1000\relax}
\let\endthebibliography=\endlist

\begin{titlepage}

\begin{center}
{\large {\bf Sensitivity of Supersymmetric Dark Matter \\ 
to the b Quark Mass}}\\
\vskip 1.5 true cm
\renewcommand{\thefootnote}
{\fnsymbol{footnote}}
 Mario E G\'omez$^{a,b}$,  Tarek Ibrahim$^{c,d}$, Pran Nath$^d$ and 
 Solveig Skadhauge$^b$  
\vskip 0.8 true cm

\noindent
{\it a. Departamento de F\'{\i}sica Aplicada, Facultad de Ciencias 
Experimentales,}\\
{\it Universidad de Huelva, 21071 Huelva, 
Spain\footnote{Current address of M.E.G.}}\\
{\it b.  Departamento de F\'\i sica, Instituto Superior T\'ecnico,
Av. Rovisco Pais, 1049-001 Lisboa, Portugal.}\\
{\it c. Department of  Physics, Faculty of Science,
University of Alexandria,} \\
{\it Alexandria, Egypt\footnote{Permanent address of T.I.}}\\ 
{\it d. Department of Physics, Northeastern University,
Boston, MA 02115-5000, USA.} \\
\end{center}

\vskip 1.0 true cm
\centerline{\bf Abstract}
\medskip
An analysis of the sensitivity of supersymmetric dark matter to 
variations in the b quark mass is given.
Specifically we study the effects on the neutralino relic 
abundance from supersymmetric loop corrections to the mass of 
the b quark. It is known that these loop corrections can 
become significant for large $\tan\beta$. 
The analysis is carried out in the 
framework of mSUGRA and we focus on the region where the relic density
constraints are satisfied  by resonant annihilation through the
s-channel Higgs poles. We extend the analysis to include CP phases 
taking into account  the mixing of the CP-even and CP-odd Higgs boson 
states which play an important role in determining the relic  
density. Implications of the analysis for the neutralino relic density
consistent with the recent WMAP relic density constraints are discussed. 
\end{titlepage}
   
\section{Introduction}    
There is now a convincing body of evidence that the universe has a 
considerable amount of non baryonic dark matter and the 
recent Wilkinson Microwave Anisotropy Probe 
(WMAP) data  allows a determination of cold  dark matter (CDM) to
lie in the range\cite{bennett,spergel}
\beqn
\Omega_{CDM} h^2 =0.1126^{+0.008}_{-0.009}
\eeqn 
One expects the Milky Way to have a similar density of cold dark matter 
and thus there are several ongoing 
experiments as well as experiments that are planned for the future 
for its detection in the laboratory\cite{dama,hdms,edelweiss,genius,cline}.
One prime CDM candidate that appears 
naturally in the framework of SUGRA models\cite{msugra} is the
neutralino\cite{goldberg}. 
We will work within the framework of SUGRA models which have a 
constrained parameter space. Thus without CP phases the mSUGRA parameter 
space is given by the parameters $m_0, m_{\frac{1}{2}}, A_0, \tan\beta$ and 
$sign(\mu)$ where $m_0$ is the universal scalar mass, 
$m_{\frac{1}{2}}$ is the universal gaugino mass, $A_0$ is the 
universal trilinear coupling (all given at the grand unification 
scale $M_G$), $\tan\beta =\langle H_2 \rangle / \langle H_1 \rangle$ 
where $H_2$ gives mass to the up quark and $H_1$ gives mass to the 
down quark and the lepton,
and $\mu$ is the Higgs mixing parameter which appears in the 
super potential in the form $\mu H_1H_2$. SUGRA models allow for
nonuniversalities and with nonuniversalities  the
parameter space of the model is enlarged. Thus, for example, 
SUGRA models with gauge kinetic energy functions that are not 
singlets of the $SU(3)\times SU(2)\times U(1)$ gauge groups 
allow for nonuniversal gaugino masses $\tilde m_i$ (i=1,2,3) at 
the grand unification scale. The parameter space of SUGRA models
is further enlarged when one allows for the CP phases. Thus
in general $\mu$, $A_0$ and $\tilde m_i$ become complex allowing for
phases $\theta_{\mu}$, $\alpha_{A_0}$, and $\xi_i$ where 
$\theta_{\mu}$ is the $\mu$ phase, $\alpha_{A_0}$ is the $A_0$ phase 
and $\xi_i$ is the phase of the gaugino mass $\tilde m_i$ ($i$=1,2,,3).
Not all the phases are independent after one performs field 
redefinitions, and only specific combinations 
of them appear in physical processes\cite{inmssm}.       

  In most of the mSUGRA parameter space the 
neutralino relic density is too large. However, there are four 
distinguishable regions where a
 neutralino relic density compatible with the WMAP constraints can 
 be found. These regions are discussed below.  
(I) The bulk region: This region corresponds to relatively small values 
of $m_0$ and $m_\frac{1}{2}$ and is dominated by sfermion exchange diagrams.
 However, it is almost ruled out by the laboratory experiments. 
(II) The Hyperbolic Branch or Focus Point region (HB/FP)\cite{ccn}: 
This region  occurs for very high 
values of $m_0$ and small values of $\mu$ and is thus close to the domain
where the electroweak symmetry breaking does not occur. 
Here the lightest neutralino has a large Higgsino component, thereby  
enhancing the annihilation cross-section to gauge boson channels. 
Furthermore,  chargino coannihilation contributes as the chargino and 
the lightest neutralino are almost degenerate. 
(III) The stau coannihilation region:  In this region
 $m_{\tilde\tau} \simeq m_\chi$ and the annihilation cross-section 
 increases due to coannihilations $\chi \tilde{\tau}_1$.  
(IV) The resonant region: 
 This is a rather broad region where the relic density constraints 
are satisfied by annihilation through resonant s-channel Higgs exchange. 
In this work we will mainly focus on, this resonant region.  

While there are many analyses of 
the neutralino relic density there are no in depth analyses of its 
sensitivity to the b quark mass. One of the purposes of this
analysis is to investigate this sensitivity. Such an analysis 
is relevant since experimentally the mass of the b quark has an
error corridor, 
and secondly because in supersymmetric theories  loop corrections  
to the b quark mass especially for large $\tan\beta$ can be
large and model dependent\cite{hall}. Recently, a full analysis of 
one loop contribution to
the bottom quark  mass ($m_b$) including phases was 
given\cite{Ibrahim:2003ca} and indeed corrections
to $m_b$  are found to be as much as 50\% or more in
some regions of the parameter space. Further, $m_b$ corrections
are found to affect considerably low  energy phenomenology
where the b quark enters\cite{Ibrahim:2003jm,Ibrahim:2003tq}.
As noted above the $m_b$  corrections are naturally large for
large $\tan\beta$ which is an interesting region because of the
possibility of Yukawa unification\cite{gomez,Baer:2001yy}
and also because it leads to large neutralino-proton  
cross-sections\cite{largetan} which makes the observation of supersymmetric
dark matter more accessible. However, we do not address the issue of Yukawa 
unification or of neutralino-proton cross-sections in this paper.
We will also discuss the dependence of the relic density on phases. 
It has been realized for some time that large phases can be accommodated
without violating the electric dipole moment (EDM) 
constraints\cite{eedm,nedm,atomic,deutron}
by a variety of ways which include mass suppression\cite{na},
the cancellation mechanism\cite{incancel,olive}, 
phases only in the third generation\cite{chang}, and other
mechanisms\cite{bdm2}. One of the important consequences  of
such phases is that the Higgs mass eigenstates are no longer
eigenstates of CP\cite{pilaftsis,inhiggs1,inhiggs2,Carena:2001fw}.
It was pointed out some time ago that CP phases 
would affect dark matter significantly in
regions where the neutralino annihilation was dominated by
the resonant Higgs annihilation\cite{inhiggs1,cin}. 
We discuss this issue in greater detail in this paper. 
        
Since the focus of this paper is on the effects of loop 
corrections to the b quark mass, we briefly 
discuss these corrections. For the b  quark the 
running mass $m_b(Q)$ and the physical mass, or the pole mass
$M_b$, are related by inclusion of QCD corrections and at 
the two loop level one has\cite{arason}
\beq 
M_b=(1+\frac{4\alpha_3(M_b)}{3\pi}+12.4\frac{\alpha_3(M_b)^2}{\pi^2})
m_b(M_b)
\eeq
where $m_b(M_b)$ is obtained from $m_b(M_Z)$ by using the 
renormalization group equations and $m_b(M_Z)$ is the running
$b$ quark mass at the scale of the $Z$ boson mass defined by  
\beq 
m_b(M_Z)=h_b(M_Z)\frac{v}{\sqrt 2}\cos\beta(1+\Delta m_b) \;.
\eeq
Here $h_b(M_Z)$ is the Yukawa coupling and $\Delta m_b$ is loop correction 
to $m_b$. Now the coupling of the b quark to the Higgs at the tree level
involves only the neutral component of the $H_1$ Higgs  boson and the 
couplings to the $H_2$ Higgs boson is absent.
However, at the loop level one finds corrections to the  
$H_1^0$ coupling as well as an additional coupling to $H_2^0$. 
Thus at the loop level the effective b quark coupling with
the Higgs is given by\cite{carena2002} 
\beqn
-L_{bbH^0}= (h_b+\delta h_b) \bar b_R b_L H_1^0 + 
\Delta h_b \bar b_R b_L H_2^0 + H.c.
\eeqn
 The correction to the b quark mass is then given 
directly in terms of $\Delta h_b$ and $\delta h_b$ so that  
\beqn
\Delta m_b= [Re(\frac{\Delta h_b}{h_b}) \tan\beta 
+Re(\frac{\delta h_b}{h_b}) ]
\eeqn
A full analysis of $\Delta m_b$ is given in Ref.\cite{Ibrahim:2003ca} and
we will use that analysis in this work.
  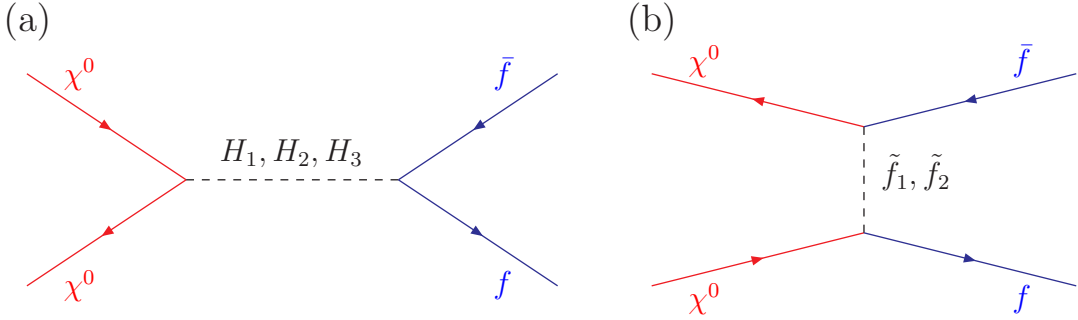
\begin{figure}
\begin{center}
\begin{picture}(200,100)(0,0)
\SetColor{Red}
\ArrowLine(60,40)(0,0)
\ArrowLine(0,80)(60,40)
\SetColor{Black}
\DashLine(60,40)(140,40){3}
\SetColor{Blue}
\ArrowLine(140,40)(200,0)
\ArrowLine(200,80)(140,40)
\Text(20,0)[]{\textcolor{red}{$\chi^{0}$}}
\Text(20,80)[]{\textcolor{red}{$\chi^{0}$}}
\Text(100,50)[]{$H_1,H_2,H_3$}
\Text(180,0)[]{\textcolor{blue}{$f$}}
\Text(180,80)[]{\textcolor{blue}{$\bar{f}$}}
\Text(0,100)[]{\large(a)}
\end{picture}
\hspace{1cm}
\begin{picture}(200,100)(0,0)
\SetColor{Red}
\ArrowLine(0,0)(80,20)
\ArrowLine(80,60)(0,80)
\SetColor{Black}
\DashLine(80,60)(80,20){3}
\SetColor{Blue}
\ArrowLine(160,80)(80,60)
\ArrowLine(80,20)(160,0)
\Text(20,-5)[]{\textcolor{red}{$\chi^{0}$}}
\Text(20,85)[]{\textcolor{red}{$\chi^{0}$}}
\Text(100,43)[]{$\tilde{f_1},\tilde{f_2}$}
\Text(140,-5)[]{\textcolor{blue}{$f$}}
\Text(140,85)[]{\textcolor{blue}{$\bar{f}$}}
\Text(0,100)[]{\large(b)}
\end{picture}
\end{center}
\caption{Feynman diagrams responsible for the main contribution to the 
neutralino annihilation cross-section in the region of the parameter
space investigated in this analysis.
Fig.(a) gives the s-channel Higgs exchange contribution and 
Fig.(b) gives the t- and u-channel sfermion 
exchange contribution to the neutralino annihilation cross-section. 
The most important decay channels for large $\tan\beta$ are for $f=b,\tau$. }
\label{ann_diags}
\end{figure}

\section{CP-even, CP-odd Higgs Mixing and b Quark Mass Corrections} 
\begin{figure}
\centering
\includegraphics[width=12cm,height=8cm]{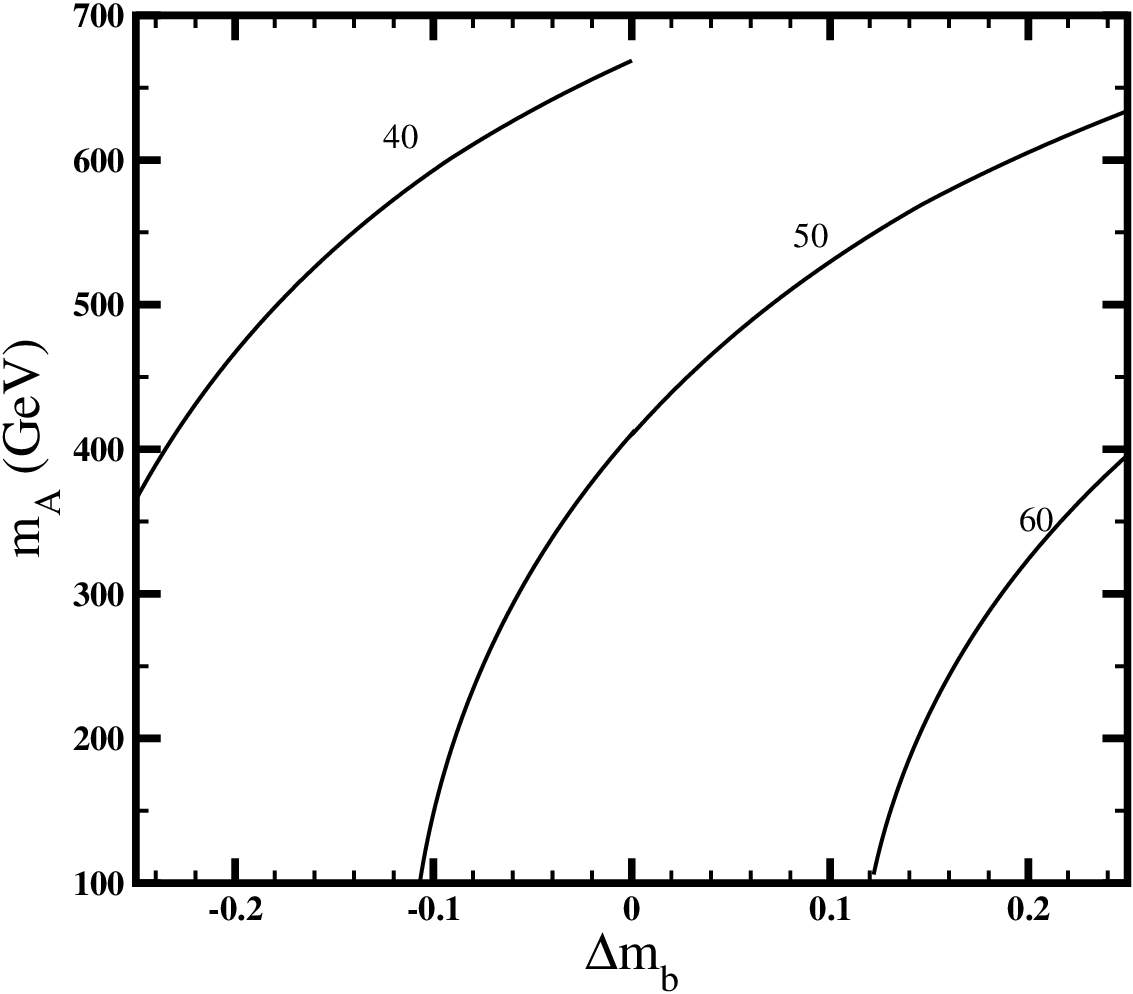}
\caption{The pseudo scalar Higgs boson mass $m_A$ as a function of
$\Delta m_b$ for fixed values of $\tan\beta$ of 40, 50 and 60
when $m_0=m_{\frac{1}{2}}=A_0=600$ GeV.}
\label{figa}
\end{figure}

As already mentioned we will focus on determining the sensitivity of 
the relic density to the b quark mass 
 in the region with resonant s-channel Higgs 
dominance. This region is characterized roughly by the constraint
\begin{equation}\label{reson_cond}
   2 m_\chi \simeq m_A  \;.
\end{equation}
The satisfaction of the relic density constraints consistent with
WMAP in this case depends sensitively on the difference
$\delta =(2 m_\chi - m_A)$ which in turn  depends sensitively on the
mSUGRA parameter space. In this context the bottom mass corrections 
are very important as the value of $m_A$ 
is strongly dependent on it, as shown in Fig.(\ref{figa}).
On the other hand at least for the domain where the neutralino is
Bino like one finds that the neutralino mass $m_\chi$ is 
rather insensitive to the  bottom mass correction and is 
almost entirely determined by $m_{\frac{1}{2}}$.
The resonant s-channel region is only open at large $\tan\beta$. 
The exact allowed range of $\tan\beta$ depends severely on 
the value of the bottom quark mass. For $\mu<0$ the resonant region 
is typically open for $\tan\beta$ in the range 35-45 and for 
$\mu>0$ for $\tan\beta$ in the range 45-55.
The large $\tan\beta$ regime is also interesting for other reasons, 
as in the presence of CP violation there can be a large mixing 
between the CP-even and the CP-odd states. Moreover, the CP phases have a 
strong impact on the b quark mass. In this section we discuss the 
relevant part of the analysis related to these effects. 
It is clear that if the CP phases influence the resonance condition,  
or equivalently the ratio $m_\chi /m_A$, they will have an impact 
on the relic density.   
This ratio is affected by phases mainly because 
 $m_A$ is strongly dependent on 
the bottom mass correction $\Delta m_b$ and through it on the CP phases. 
Furthermore, the Higgs couplings relevant for computing the 
annihilation cross-section depend on the CP phases. 
Thus we expect the relic density to be strongly 
dependent on the CP phases. 

We begin by considering  the s-channel decay to a pair of fermions, as 
shown in Fig.(\ref{ann_diags})(a).
The Yukawa coupling correction enters clearly here
in the vertex of the neutral Higgs with the fermion pair.
The amplitude for $\chi(p_1)\chi(p_2)\rightarrow f(k_1)\bar{f}(k_2)$,  
mediated by Higgs mass eigenstates, $H_k,\; k=1,2,3$ may be 
written as, 
\begin{equation}
M_k^f = \bar{v}(p_2)\left[S_k'- i S_k'' \gamma_5\right]u(p_1)\frac{1}
{-M_{H_k}^2+s-i M_{H_k}\Gamma_{H_k}}
\bar{u}(k_1)\left[C^S_{f,k}+i C^P_{f,k} \gamma_5\right]v(k_2)
\label{cscp}
\end{equation}
where
\beq
S'_k=\frac{g m_{\chi}R_{k2}}{2 M_w \sin\beta}+Re(A_k),
\eeq
\beq
S''_k=-\frac{g m_{\chi}R_{k3}\cot\beta}{2 M_w}+Im(A_k).
\eeq
and the parameters $A_k$ are defined by
\beq
A_k=Q''^{*}_{00}  R_{k1} - i Q''^{*}_{00}  R_{k3}\sin\beta
-\frac{1}{\sin\beta}(R_{k2} - i R_{k3}\cos\beta)(Q''^{*}_{00} 
\cos\beta +R''^{*}_{00} ),
\eeq
where
\beq
Q''_{00} =X_{30} ^{*}(X_{20} ^{*} -\tan\theta_W X_{10} ^{*})
\eeq
and
\beq
R''_{00} = \frac{1}{2 M_w} [\tilde{m}_2 X^{*2}_{20} + \tilde{m}_1 X^{*2}_{10}
-2 \mu^{*} X_{30}^{*} X_{40}^{*}].
\eeq
Here $X$ is the matrix that diagonalizes  the neutralino mass
matrix so that $X^TM_{\chi}X$ = $diag (m_{\chi_1}, m_{\chi_2},
m_{\chi_3}, m_{\chi_4})$  and $m_{\chi_0}$ is the lightest
neutralino.  Thus 0 is the index among $1,2,3,4$ that corresponds 
to the lightest neutralino (later in the analysis we will drop 
the subscript on $\chi_0$ and $\chi$ will stand for the lightest
neutralino).  

Since the CP effects in the Higgs sector play an important
role in this analysis, we briefly review the main aspects of this phenomena.
In the presence of explicit CP violation the two Higgs doublets of
the supersymmetric standard  model (MSSM) can be decomposed as follows
\beqn
(H_1)= \left(\matrix{H_1^0\cr
 H_1^-}\right)
 =\frac{1}{\sqrt 2} 
\left(\matrix{v_1+\phi_1+i\psi_1\cr
             H_1^{-'}}\right)\nonumber\\
(H_2)= \left(\matrix{H_2^+\cr
             H_2^0}\right)
=\frac{e^{i\theta_H}}{\sqrt 2} \left(\matrix{H_2^{+'} \cr
             v_2+\phi_2+i\psi_2}\right)
\eeqn
where  $\phi_1, \phi_2, \psi_1,\psi_2$ are real quantum fields 
and $\theta_H$ is a phase. Variations  with respect to the fields
give 
\beqn
-\frac{1}{v_1}(\frac{\partial \Delta V}{\partial \phi_1})_0=
m_1^2+\frac{g_2^2+g_Y^2}{8}(v_1^2-v_2^2)+m_3^2 \tan\beta \cos\theta_H
\nonumber\\
-\frac{1}{v_2}(\frac{\partial \Delta V}{\partial \phi_2})_0=
m_2^2-\frac{g_2^2+g_Y^2}{8}(v_1^2-v_2^2)+m_3^2 \cot\beta \cos\theta_H
\nonumber\\
\frac{1}{v_1}(\frac{\partial \Delta V}{\partial \psi_2})_0=
m_3^2 \sin\theta_H= \frac{1}{v_2}
(\frac{\partial \Delta V}{\partial \psi_1})_0
\label{minimization}
\eeqn
where $m_1, m_2, m_3$ are the parameters that enter in the 
tree-level Higgs potential, i.e.,
$V_0=$ $m_1^2 |H_1|^2+ m_2^2 |H_2|^2+(m_3^2 H_1.H_2 + H.c.) + V_D$ where 
$V_D$ is the D-term contribution,  $g_2$ and $g_Y$ are the gauge coupling
constants for $SU(2)$ and $U(1)_Y$ gauge groups, and 
$\Delta V$ is the loop correction to the Higgs potential.  
In the above the subscript 0 denotes that the quantities are 
computed at the point where   $\phi_1=\phi_2=\psi_1=\psi_2=0$.
Eq.(\ref{minimization}) provides a determination of $\theta_H$.
Computations in the above basis lead to a $4\times 4$ Higgs mass  
matrix.  It is useful to introduce a new basis 
$\{ \phi_1,\phi_2,\psi_{1D}, \psi_{2D}\}$ where 
 $\psi_{1D},\psi_{2D}$ are defined by
\beqn
\psi_{1D}=\sin\beta \psi_1+ \cos\beta \psi_2\nonumber\\
\psi_{2D}=-\cos\beta \psi_1+\sin\beta \psi_2
\eeqn
In the new basis the field $\psi_{2D}$ exhibits itself as the
Goldstone field and decouples from the other three
fields $\{ \phi_1,\phi_2,\psi_{1D}\}$ and the Higgs mass matrix
in the new basis takes on the form
\beq
M^2_{\rm Higgs}=
\left(\matrix{M_Z^2c_{\beta}^2+m_A^2s_{\beta}^2+\Delta_{11} &
-(M_Z^2+m_A^2)s_{\beta}c_{\beta}+\Delta_{12} &\Delta_{13}\cr
-(M_Z^2+m_A^2)s_{\beta}c_{\beta}+\Delta_{12} &
M_Z^2s_{\beta}^2+m_A^2c_{\beta}^2+\Delta_{22} & \Delta_{23} \cr
\Delta_{13} & \Delta_{23} &(m_A^2+\Delta_{33})}\right)
\label{mass2higgs}
\eeq
where $m_A$ is the mass of the CP-odd Higgs boson at the tree level,
$M_Z$ is the Z boson mass, $s_{\beta}(c_{\beta})=\sin\beta (\cos\beta)$,
and $\Delta_{ij}$ are the loop corrections. These loop corrections 
have been computed from the exchange of stops and sbottoms in 
Refs.\cite{pilaftsis,inhiggs1}, from the exchange of
charginos in Ref.\cite{inhiggs1} and from the exchange of
neutralinos in Ref.\cite{inhiggs2}.
Thus the corrections $\Delta_{ij}$ (i,j=1,2,3) receive contributions 
from stop, chargino and neutralino exchanges. Their relative contributions
depend on the point in the parameter space one is in. 
We denote the eigenstates of the mass$^2$ matrix of Eq.(\ref{mass2higgs}) 
by $H_k$ (k=1,2,3)
and we  define the matrix $R$ with elements $R_{ij}$  as the matrix
which diagonalizes the above $3 \times 3$ Higgs mass$^2$ matrix 
so that 
\beq
R M^2_{\rm Higgs} R^T =diag(M^2_{H_1},M^2_{H_2},M^2_{H_3}).
\eeq
and thus we have 
\beq
\left(\matrix{H_{1} \cr 
H_{2} \cr 
H_{3}} \right)= R \times 
\left(\matrix{
\phi_{1} \cr 
\phi_{2} \cr 
\psi_{1D}}\right).
\eeq
In the analysis of this paper we work in the decoupling regime of
the Higgs sector,  characterized by 
$m_A \gg M_Z$ and large $\tan\beta$.
In this regime the light Higgs boson is denoted by $H_2$ and the  
 two heavy Higgs particles are described by $H_1$ and $H_3$.
For the case when we have CP conservation and no mixing of CP even
and CP odd states, we denote the heavy scalar Higgs boson by $H$ 
(at large $\tan \beta$ it is almost equal to $\phi_1$) and the 
pseudo scalar  Higgs boson by $A$. Returning to the general case
with CP phases, in the decoupling limit the heavy Higgs states are 
almost degenerate and moreover have nearly equal widths, i.e.,   
\begin{equation}\label{decoup}
m_{H_1} \simeq m_{H_3}\,,\qquad \Gamma_{H_1} \simeq \Gamma_{H_3} \,.
\end{equation}
Furthermore, the lightest Higgs boson behaves almost like the SM Higgs 
particle.  This means that there may be considerable mixing between 
the two heavy CP eigenstates, $H$ and $A$, whereas the mixing with 
the lightest Higgs is tiny. 
Corrections to Yukawa coupling arise through the parameters 
$C_{q,k}^{S,P}$ that enter in  Eq.(\ref{cscp}) so that 
\beq
C_{b,k}^S = \bar{C}_{b,k}^S \cos \chi_b -\bar{C}_{b,k}^P \sin \chi_b,
\eeq
and
\beq
C_{b,k}^P = \bar{C}_{b,k}^S \sin \chi_b +\bar{C}_{b,k}^P \cos \chi_b,
\eeq
where 
\beq
\sqrt{2} \bar{C}_{b,k}^S = Re (h_b +\delta h_b) R_{k1} + [-Im (h_b +\delta h_b)
\sin\beta + Im (\Delta h_b) \cos \beta] R_{k3} + Re (\Delta h_b) R_{k2}
\eeq    
and where 
\beq
\sqrt{2} \bar{C}_{b,k}^P =- Im (h_b +\delta h_b) R_{k1} + 
[-Re (h_b +\delta h_b)
\sin\beta + Re (\Delta h_b) \cos \beta] R_{k3} - Im (\Delta h_b) R_{k2}
\eeq
and the angle $\chi_b$ is defined by 
\beq
\tan \chi_b =\frac{Im (\frac{\delta h_b}{h_b}+\frac{\Delta h_b}{h_b}
\tan\beta)}{1+ Re(\frac{\delta h_b}{h_b}+\frac{\Delta h_b}{h_b}
\tan\beta)}
\eeq
The phases enter in a variety of ways in the model.
Thus the parameters  $Q''_{00}$ and $R''_{00}$ contain the combined 
effects of the phases $\theta_{\mu}$, $\xi_1$ and $\xi_2$.
Similarly, $R_{ij}$ contain the combined effects of the
above three phases and in addition depend on $\alpha_{A_f}$
(of which the most important is $\alpha_{A_t}$). 
Further, $C^{S,P}$ derive their phase dependence through $R_{ij}$ and
in addition depend on $\xi_3$ which enters via the 
SUSY QCD corrections $\Delta h_b$ and $\delta h_b$.
Including all the contributions any of the phases may produce
a strong effect on the relic density. Explicit analyses bear this
out although the relative contribution of the different phases
depends on the part of the parameter space one is working in.
The $s$--channel annihilation cross-section for 
$\chi(p_1)\chi(p_2)\rightarrow f(k_1)\bar{f}(k_2)$  
is proportional to the squared of the amplitude given in Eq.(\ref{cscp}) 
and reads
\begin{equation}\label{annsigma}
M_k^f (M_l^f)^*=\frac{(C^S_{f,k} C^S_{f,l}+C^P_{f,k}C^P_{f,l})
\left[S_k' S_l' (1-4 m_\chi^2/s)+S_k'' S_l''\right]}
{(-M_{H_k}^2+s-i M_{H_k}\Gamma_{H_k})(-M_{H_l}^2+s+i M_{H_l}\Gamma_{H_l})}s^2 
\end{equation}
Therefore, the imaginary couplings, $S_k''$, will yield the dominant 
contribution to the thermally averaged annihilation cross-section, 
as the real couplings, $S_k'$, are p-wave suppressed by the 
factor $(1-4 m_\chi^2/s)$. 
In the case of vanishing CP-phases the  pseudo scalar  mediated channel 
thereby dominates over the one mediated by the heavy scalar Higgs. 
However, the contribution from 
 $H$ mediation cannot be neglected, as its contribution is 
typically about 10\%.
In the presence of non-zero phases both of the heavy Higgs    
acquire imaginary coupling and both may give a significant 
contribution. We may neglect the contribution from the 
lightest Higgs exchange diagram since it is not 
resonant and moreover is suppressed by small 
couplings\footnote{The region where the lightest Higgs is 
resonant is almost excluded by laboratory constraints.}.

\begin{figure}
\centering
\includegraphics[width=12cm,height=8cm]{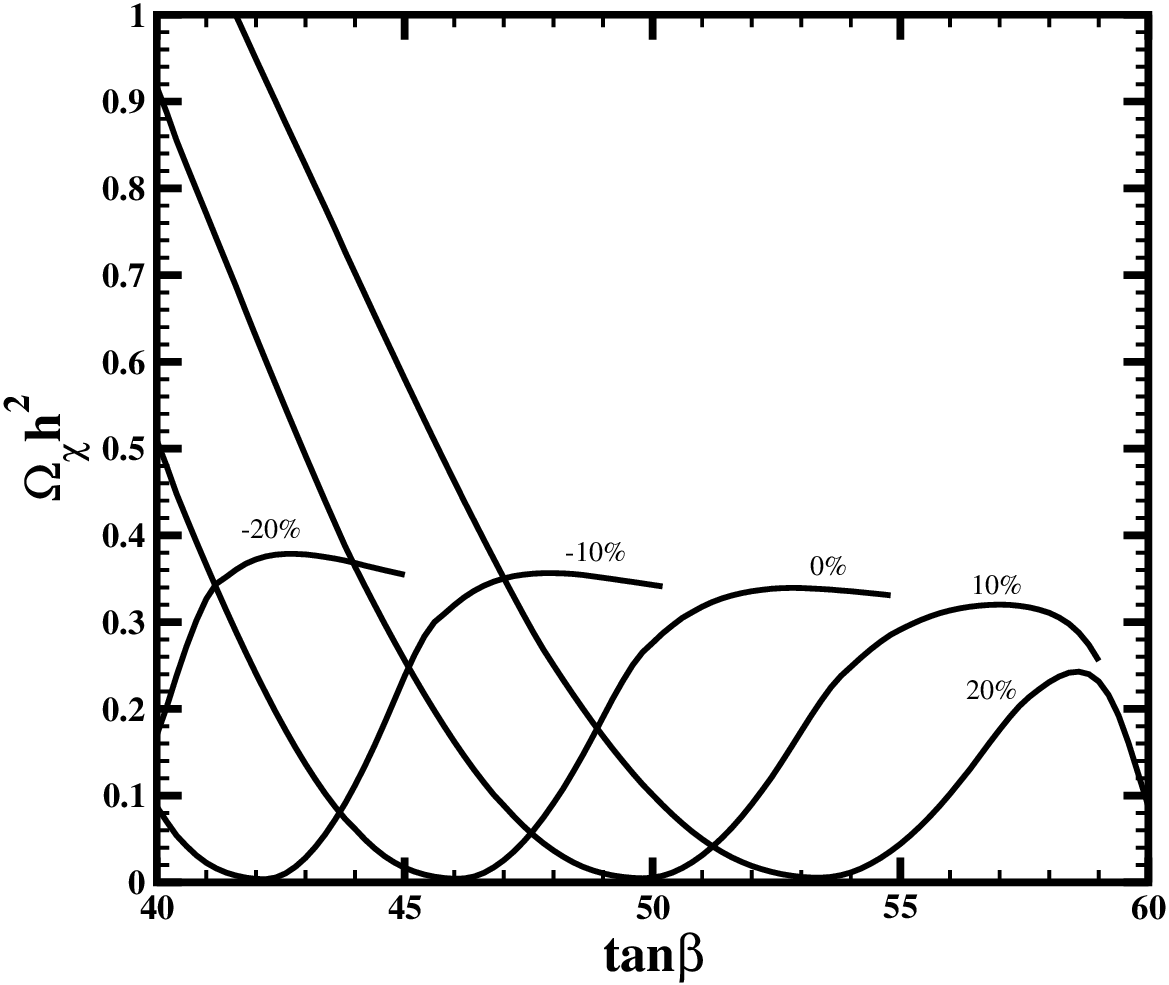}
\caption{An exhibition of the sensitivity of the relic density
to the b quark mass as a function of $\tan\beta$ for the case
when $m_0=m_{\frac{1}{2}}=A_0=600$ GeV for values of 
$\Delta {m_b}$ from $(-20\%) -(+20\%)$.}
\label{ohtan}
\end{figure}
As already mentioned the inclusion of the CP phases has two major 
consequences; it affects the SUSY correction to the bottom mass $\Delta m_b$ 
and it also generates a mixing in the heavy Higgs sector.  
We discuss now in greater detail the effect of mixing in the heavy
 Higgs sector. We begin by observing that in 
 the CP conserving case the pseudo scalar  channel 
gives the main contribution. As the Higgs mixing turns 
on the pseudo scalar  becomes a linear combination of the two mass eigenstates 
$H_1,H_3$, whereas $H_2$ stays almost entirely a CP-even state. 
However, the total annihilation cross-section which is a sum over 
 all the  Higgs exchanges remains almost constant.  Since CP even and
 CP odd Higgs mixing involves essentially only two Higgs bosons, 
 we may represent this mixing by just one $2\times 2$ orthogonal matrix 
 rotation. Such a rotation  does not change the sum of 
the squared couplings of the two heavy Higgs boson, and thereby 
the effect of the mixing on the annihilation cross-section is small.
The basic reason for the mixing effect being small is because of
the near degeneracy of the CP even and CP odd Higgs masses and
widths, i.e., 
the fact that $m_{H_1}\approx m_{H_3}$, and $\Gamma_{H_1}\approx \Gamma_{H_3}$.
We note in passing that the 
 contribution from the Higgs exchange interference term 
 $H_h-H_k$ to the neutralino 
 annihilation cross section is negligible. These phenomena allow
  us to write the total s-channel 
contribution in a simplified way. Thus recalling that the lightest Higgs 
gives almost a vanishing contributions, we only have to sum 
over the heavy Higgs particles in Eq.(\ref{annsigma}). 
As we are in the decoupling limit given  by Eq.(\ref{decoup}), 
the propagators in Eq.(\ref{annsigma}) are identical and can be 
factored out.  
Furthermore, for large $\tan\beta$  we have the 
approximate relations between the bottom-Higgs couplings 
in the CP-conserving case,
\begin{equation}\label{coup_cpc}
 C^S_{\phi_1} \simeq -C^P_{A} ,  \qquad   C^S_{A} \simeq  C^P_{\phi_1} \;.
\end{equation}
These relations are independent of rotations in the Higgs sector, i.e., 
Higgs mixing, 
as is easily checked. Also because of the decoupling of the light
Higgs boson, the mixing of the Higgs is described by just one angle
so that 
\begin{equation}
\left(\matrix{  H_1 \cr H_3 \cr} \right) = 
\left(\matrix{ \cos(\theta) & \sin(\theta) \cr -\sin(\theta) & \cos(\theta)
\cr} \right) 
\left(\matrix{ H \cr  A \cr} \right)
\end{equation}
which gives: 

\begin{eqnarray}
 C^S_{b,1} = \cos(\theta)C^S_{\phi_1} + \sin(\theta) C^S_A \\
 C^S_{b,3} = -\sin(\theta)C^S_{\phi_1} + \cos(\theta) C^S_A
\end{eqnarray}
and 
\begin{eqnarray}\label{b_coup_rel}
 C^P_{b,1} = \cos(\theta)C^P_{\phi_1} + \sin(\theta) C^P_A
=  \cos(\theta)C^S_A- \sin(\theta)C^S_{\phi_1}=C_{b,3}^S \\
 C^P_{b,3} = -\sin(\theta)C^P_{\phi_1} + \cos(\theta) C^P_A
= -\sin(\theta) C^S_A -  \cos(\theta) C^S_{\phi_1} = - C^S_{b,1} 
\end{eqnarray}
This is just Eq.(\ref{coup_cpc}) in the Higgs rotated basis 
and we see that the interference terms are very small
\begin{equation}
C^S_{b,1} C^S_{b,3} + C^P_{b,1} C^P_{b,3} \simeq 0 \;.
\end{equation}
Furthermore, using Eq.(\ref{annsigma}) it is clear 
that the s-channel contribution is proportional to $C_s$ where 
\begin{equation}
 C_s = \left( (C^S_{b,1})^2 + (C^P_{b,1})^2 \right) S_{1}''^2 
+ \left( (C^S_{b,3})^2 + (C^P_{b,3})^2 \right) S_{3}''^2
\end{equation}
The b quark couplings factors out, due to  Eq.(\ref{b_coup_rel}), 
and we get,
\begin{equation}
 C_s = \left( (C^S_{b,3})^2 + (C^P_{b,3})^2 \right) 
(S_{1}''^2 + S_{3}''^2) \,.
\end{equation}
Again, the Higgs mixing does not change the square of the imaginary 
Higgs-$\chi$-$\chi$ coupling. In the CP conserving limit we get 
\begin{equation}
 S_{\phi_1}''=0 , \qquad S_{A}'' = -\frac{g m_\chi \cot(\beta)}{2 M_W}
-Q_{00}''\sin(\beta) + \cot(\beta)(Q_{00}''\cos(\beta)+R_{00}'')
\end{equation}
and 
\begin{equation}
(S_{1}''^2 + S_{3}''^2) \rightarrow 
(\sin(\theta) S_{A}'')^2 + (\cos(\theta) S_{A}'')^2 = (S_{A}'')^2  
\end{equation}
Thus $C_S$ is unaffected by the Higgs mixing, but can 
vary with phases if the magnitudes $|S_{A}''|$ and 
$|C^S_{b,3}|^2 + |C^P_{b,3}|^2$  vary  with phases. 
As already discussed the CP phases have a large impact on the 
relic density through their influence on the b quark mass via 
the loop correction $\Delta m_b$. 
An exhaustive analysis of the dependence of $\Delta m_b$ on phases
is given in Ref.\cite{Ibrahim:2003ca}.  
For large $\tan\beta$ and small $A_0$ 
the dominant contribution to $\Delta m_b$ comes from the 
gluino-sbottom exchange diagram and the important phases here are
$\theta_{\mu}$ and $\xi_3$. However, if $A_0$ is  
large the stop-chargino correction  would be large and
 the phase $\alpha_{A_t}$ plays an important role.
There are also neutralino diagrams but normally their 
contributions are small.   
Thus, the CP phases  $\theta_\mu, \xi_3$ and  $\alpha_{A_0}$ 
may strongly affect the relic density, 
whereas only weak dependent on  $\xi_1, \xi_2$ will be present.    
\begin{figure}
\hspace*{-0.6in}
\centering
\includegraphics[width=12cm,height=8cm]{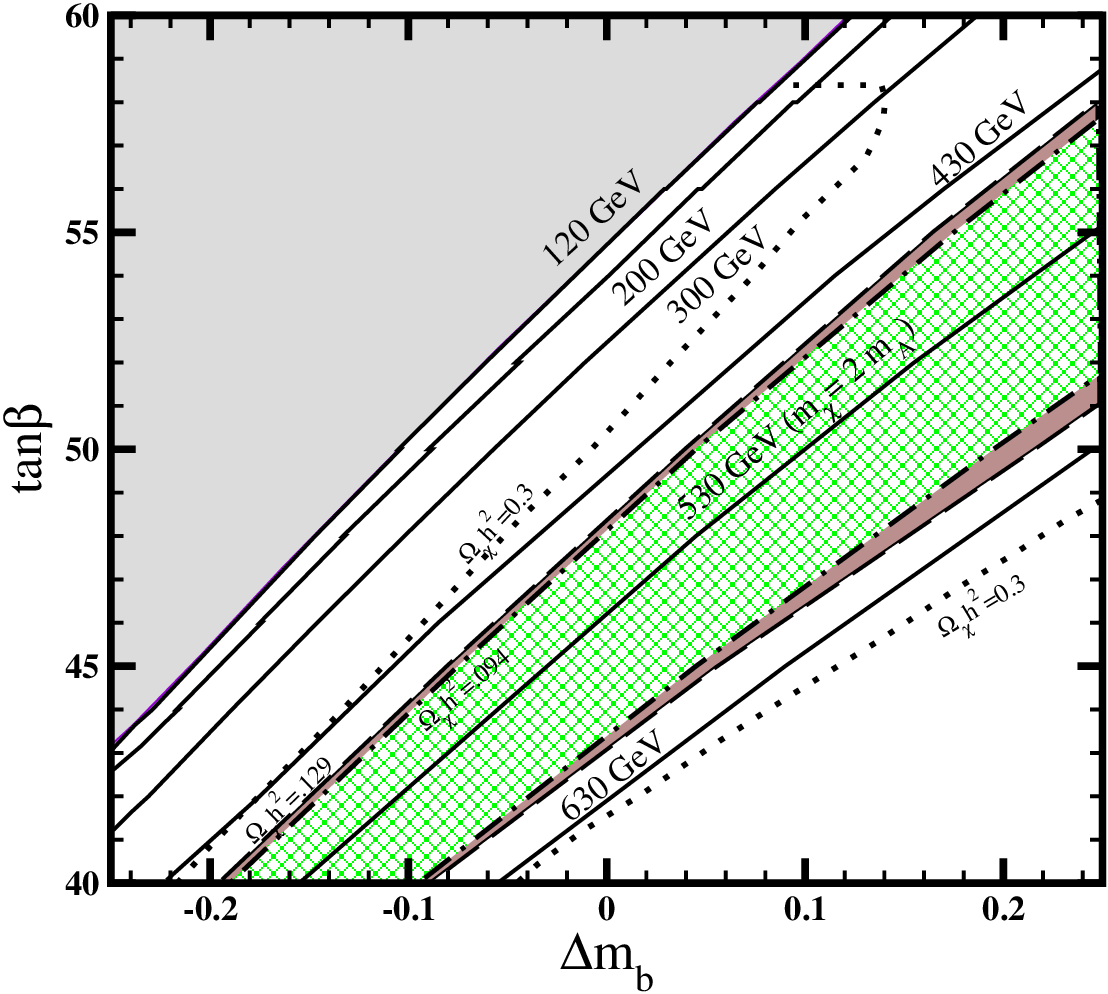}
\caption{The region allowed by the relic density
constraints in the $\tan\beta$ - $\Delta m_b$ plane for the case
when $m_0= m_{\frac{1}{2}}=A_0=600$ GeV. Curves with fixed $m_A$ 
are also shown.} 
\label{tandel}
\end{figure}

\section{Sensitivity of Dark Matter to the b Quark Mass
without CP Phases}
While a considerable body of work already exists on the analyses of
supersymmetric dark matter (for a small sample see Ref.(\cite{direct})),
no in depth study exists on the sensitivity of  
dark matter analyses to the  b quark mass. In this section we
analyse this sensitivity of the relic density to the b quark mass
for the case when the phases  are  set to zero.
In the analysis we use the standard techniques of evolving
the parameters given in mSUGRA at the grand unification 
scale by the renormalization group evolution taking care that
charge and color conservation is appropriately preserved
(for a recent analysis of charge and color conservation 
constraints see Ref.\cite{gomez1}).
We describe now the result of the analysis. 
(For a partial previous analysis of this topic see 
Ref\cite{Ellis:2003si}). One of the parameters
which enters sensitively in the dark matter analysis is  the mass 
of the CP odd Higgs boson $m_A$. Fig.~(\ref{figa}) shows 
$m_A$ as a function of the b quark correction $\Delta m_b$, which is 
used as a free parameter.
The ranges chosen are such that the $m_A$ may lie in the
resonance region of the annihilation of the two neutralinos.
We find that $m_A$ shows a very significant variation as 
$\Delta m_b$ moves in the range $-.3$ to $.3$. Fig.~(\ref{figa}) 
demonstrates the huge sensitivity of $m_A$ to the b quark mass. 
Fig.~\ref{figa} also shows that for fixed $\tan\beta$ one can 
enter in the area of the resonance for certain values of $\Delta m_b$. 
Fig.~\ref{ohtan} shows the sensitivity of the relic density to  
corrections to the b quark mass.
The analysis was carried out using {\it Micromegas}\cite{micromegas}. The 
dominant channels that contribute to the relic density depend
on the mass region and are as follows: In the 
region  $2 m_\chi\ll m_A$ the main channels are 
$\chi\chi \rightarrow \tau\bar{\tau}$ and  
$\chi\chi \rightarrow b\bar{b}$. 
Here typically $\Omega_{\chi} h^2 > 0.5$ and the main contribution comes from  
t- and u-channel exchange of the sbottom and stau sparticles. 
Moreover, also the effects of the $\mu$ and $e$ decay channels can be seen. 
Since their contributions are suppressed by the corresponding 
slepton masses, it signifies that one is far away from the 
s-channel Higgs resonances.
In the region $2 m_\chi\sim m_A$ the resonant channels account for 
almost the full contribution to $\Omega_{\chi} h^2$ and their influence can be 
detected several widths, $\Gamma_A$, away from the resonance. In this 
region the contribution to the neutralino relic density from 
the t- and u-channel exchanges can be as much as 10\%  within the
relic density range allowed by the WMAP data.

Another contribution that can potentially enter is coannihilation.
Indeed for $m_{\tilde{\tau}_1} < 1.25\times m_\chi$, one
has important effects from  $\tilde{\tau}_1 \chi$ coannihilation. 
These effects can be observed at the end of the lines of 
$\Delta m_b=10\%, 20\%$ in Fig.~(\ref{ohtan}).
Thus for low values of $\tan\beta$ one is in the non resonant region
and increasing $\tan\beta$ moves one to the resonant region and
consequently the relic  density decreases due to resonant
annihilation. 
As $\tan\beta$ increases further, the relic density 
increases to become flat due to the non fermionic decays. 
Finally, the curves for $\Delta m_b$ of $10\%$ and $20\%$
exhibit coannihilation and  $\Omega_{\chi} h^2$ decreases again due to 
this effect as $\tan\beta$ increases.   
\begin{figure}[t]
\hspace*{-0.6in}
\centering
\includegraphics[width=8cm,height=9cm]{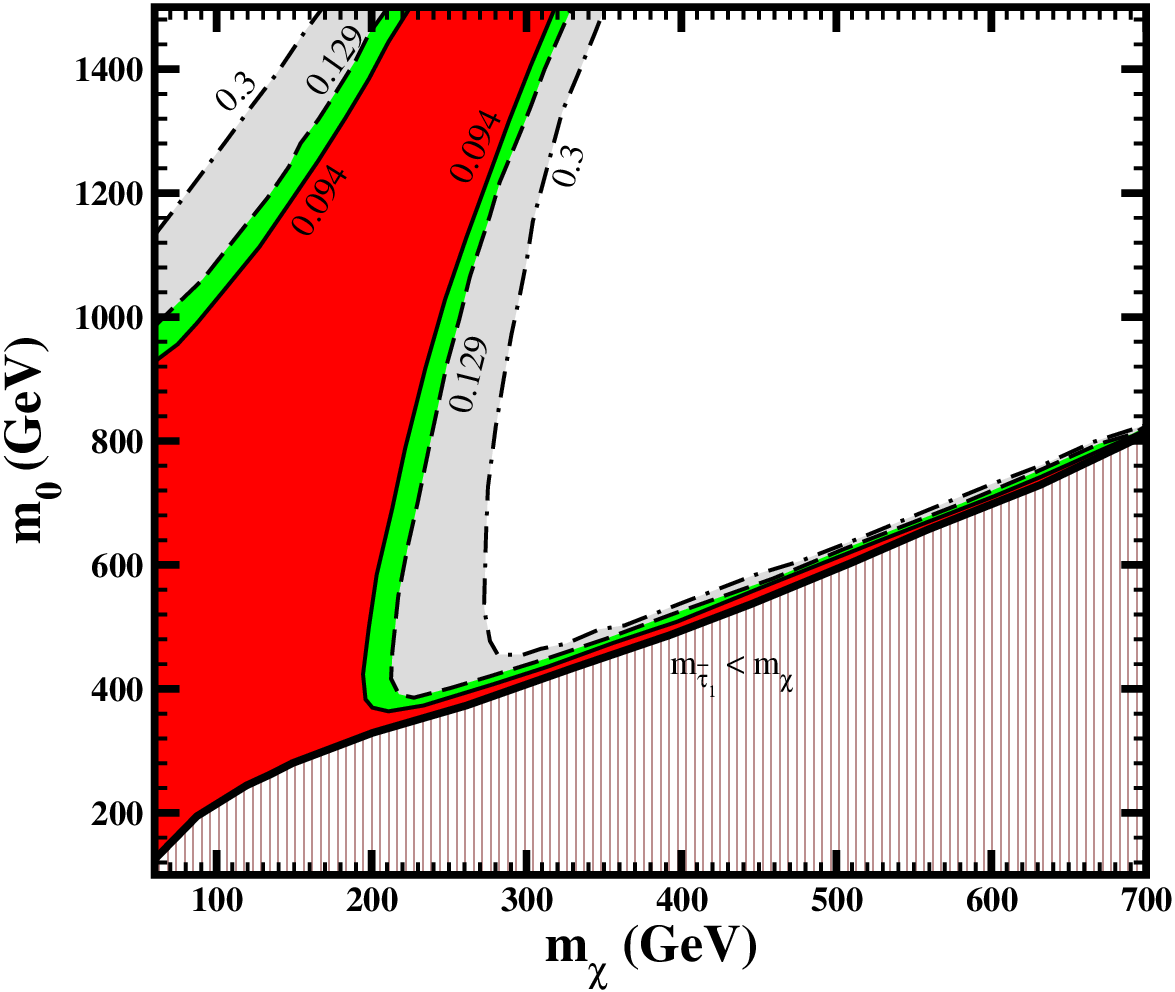}
\includegraphics[width=8cm,height=9cm]{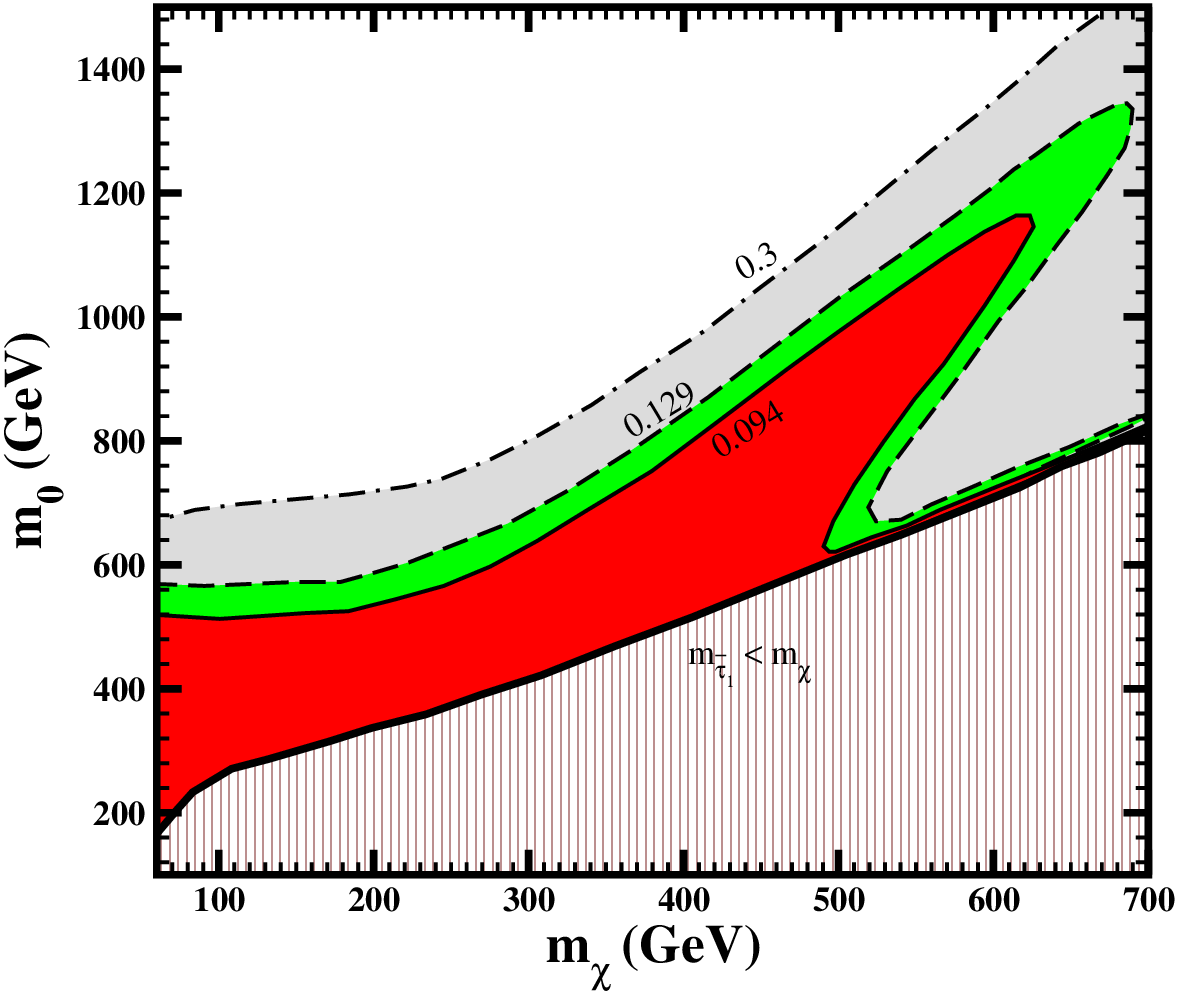}
\caption{Region in the $m_0-m_{\chi}$ plane of mSUGRA allowed by the relic 
density constraints for the case when $\tan\beta =50$, $|A_0|=m_{\frac{1}{2}}$ 
for values of $\Delta m_b=0$ (left) and  $\Delta m_b=20\%$ (right). The 
limiting lines close areas such that $\Omega_{\chi} h^2$ is below the 
indicated value.}
\label{m0lsp1}
\end{figure}
In Fig.(\ref{tandel}) regions with fixed corridors of 
the relic density are plotted in the $\tan\beta-\Delta m_b$ plane.  
The region consistent with the current range of relic density
observed by WMAP is displayed in the dark shaded region. 
The hatched region has a value of $\Omega_{\chi} h^2$ below the 
WMAP observation. 
Furthermore, curves with constant values of $m_A$ are exhibited. 
The analysis shows that the region consistent with the WMAP relic 
density constraint is very sensitive to the $\Delta m_b$ correction.

Fig.(\ref{m0lsp1}) displays area plots in the $m_0-m_\chi$
plane of the relic density. The light shaded/grey region 
has $0.1291<\Omega_{\chi} h^2<0.3$. The medium shaded/green region has 
$0.094<\Omega_{\chi} h^2<0.1291$ and is the $2\sigma$ allowed WMAP region. 
Finally the dark shaded/red region has $\Omega_{\chi} h^2<0.094$. 
The hatched region is excluded as the $\tilde\tau$ is the LSP.
In Fig.(\ref{m0lsp1}) we considered
two values of $\Delta m_b$;  $\Delta m_b=0$ and 
$\Delta m_b=20\%$. A comparison of these two exhibits the 
dramatic dependence of the various regions on the b quark mass.
Specifically it is seen that the region consistent with the WMAP
constraint is drastically shifted toward lower values of $m_\chi/m_0$ for 
smaller $\Delta m_b$. 
 Inclusion of 
the experimental bounds from processes such as 
$b\rightarrow s \gamma$ on Fig.~(\ref{m0lsp1}) is beyond the scope of our 
study. However, the case of  $\Delta m_b=20\%$ is comparable to the 
mSUGRA case (where  $\Delta m_b$ ranges approximately from 17 to 21\%). 
Therefore, the restrictions from the bounds on 
$b\rightarrow s \gamma$, along with other constraints, on the left graph 
of  Fig.~(\ref{m0lsp1}) can 
be approximately deduced from the appropriate figures of Ref.~\cite{gomez1}.
\begin{figure}
\centering
\includegraphics[width=12cm,height=8cm]{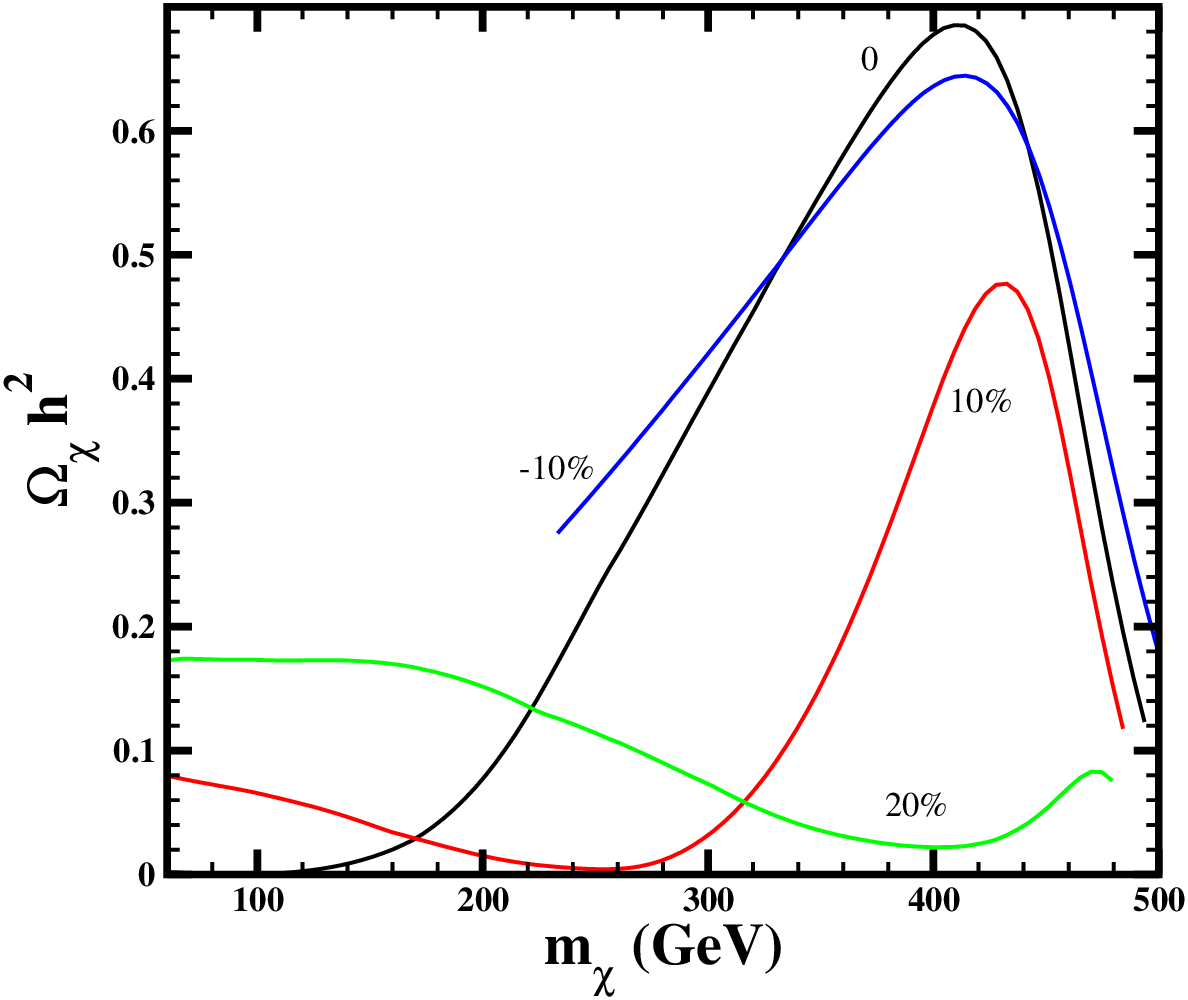}
\caption{Plot of $\Omega_{\chi} h^2$ as a function of $m_{\chi}$ 
for values of the $\Delta m_b$ varying from $(-10\%) - (+20\%)$ 
when $m_0=600$ GeV, $m_{\frac{1}{2}}=|A_0|$ and $\tan\beta=50$.}
\label{ohlsp}
\end{figure}
We note that in Fig.(\ref{m0lsp1}) there is a region where $m_0$
and $m_{\frac{1}{2}}$ get large and appear to have a superficial 
resemblance with the Hyperbolic Branch/Focus Point (HB/FP) 
region\cite{ccn}.  However, the mechanism by which relic density
constraint is satisfied in the WMAP region is entirely different in this
case than in the HB/FP case\cite{wmap1,wmap2}. Thus in the analysis
presented here the relic density constraints are satisfied by
the mechanism of proximity to a resonant state (see also in 
this context Ref.\cite{baerhiggspole}) while for the HB/FP region
the satisfaction occurs with a significant amount of coannihilation.
In Fig.(\ref{ohlsp}) we give a plot of $\Omega_{\chi} h^2$ as a 
function of $m_{\chi}$ for fixed $m_0$ (i.e., $m_0=600$ GeV)
and $\tan\beta =50$ for $\Delta m_b$ values varying in the range 
$(-10\%)-(+20\%)$. Again one finds that the relic density is sharply
dependent on the b quark mass correction.

\section{Sensitivity of Dark Matter to the b Quark Mass with CP Phases}    
\begin{figure}[t]
\hspace*{-0.6in}
\centering
\includegraphics[width=8cm,height=9cm]{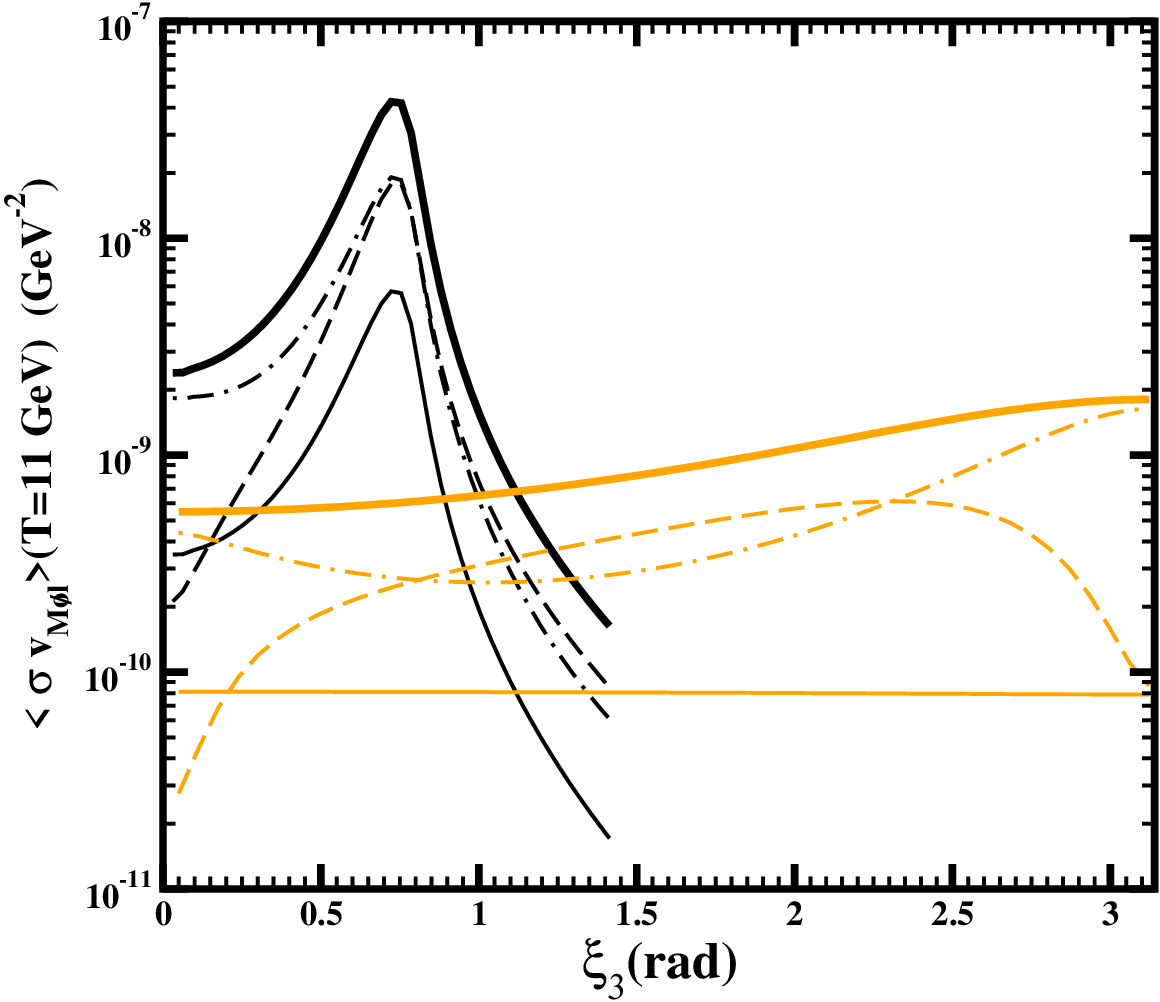}
\includegraphics[width=8cm,height=9cm]{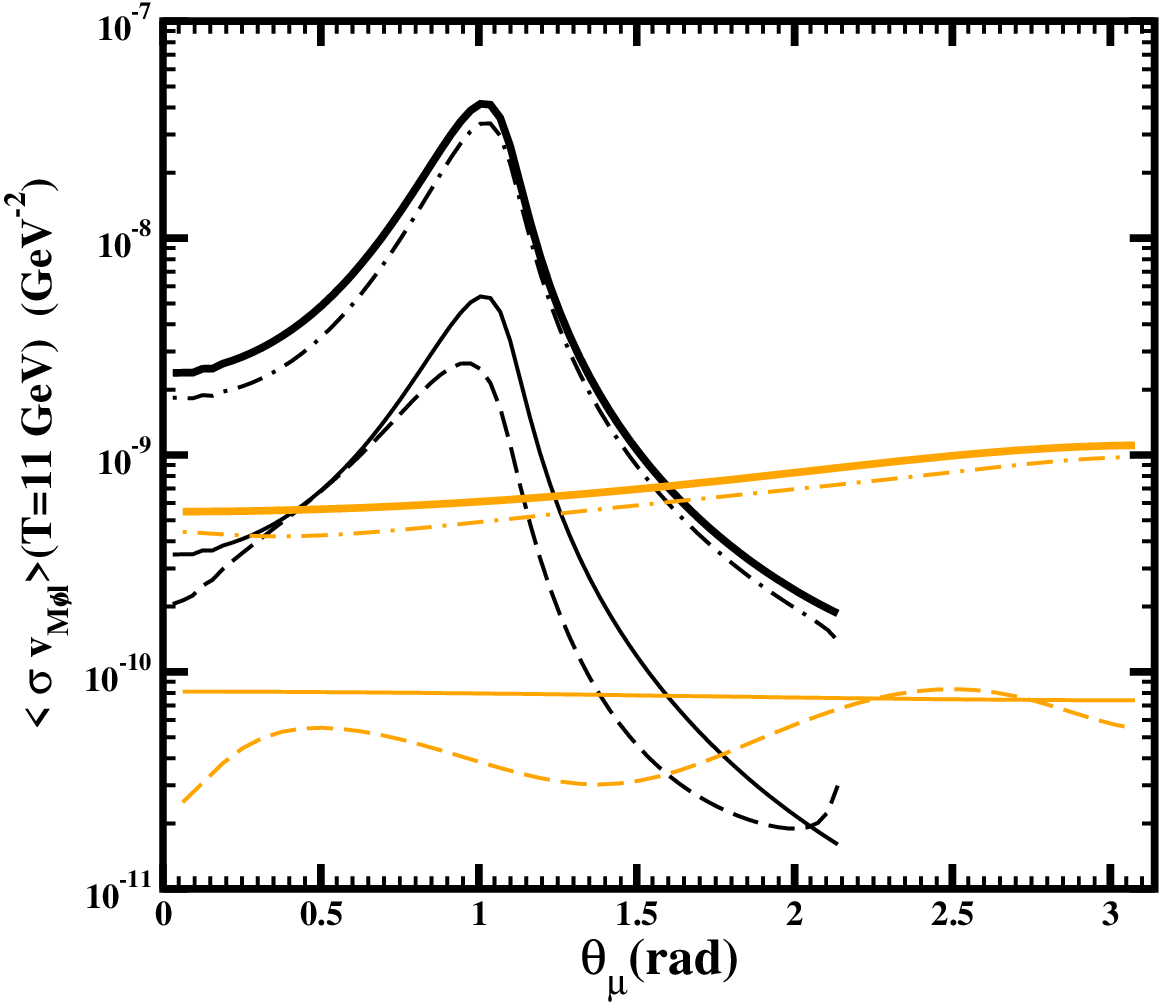}
\caption{A plot of $\langle \sigma \vmol \rangle$ 
as a function of $\xi_3$ and $\theta_{\mu}$ 
(with the other phases set to zero) for the case 
when $m_0=m_{\frac{1}{2}}=A_0=600$ GeV, $\tan\beta =50$ and 
using the theoretically predicted value of $\Delta m_b$ (black lines), 
$\Delta m_b=0$ (light lines). 
The contribution of dominant channels to 
$\langle \sigma \vmol \rangle$ are also shown: all 
contributions (thick lines), only s-channel $H_1$ mediated annihilation 
to $b\bar{b}$ (dashed lines) and only s-channel $H_3$ mediated annihilation 
to $b\bar{b}$ (dot-dashed lines) and all s-channel annihilation 
to $\tau \bar{\tau}$ (solid thin lines).}
\label{crtan50}
\end{figure}
\begin{figure}[t]
\hspace*{-0.6in}
\centering
\includegraphics[width=8cm,height=9cm]{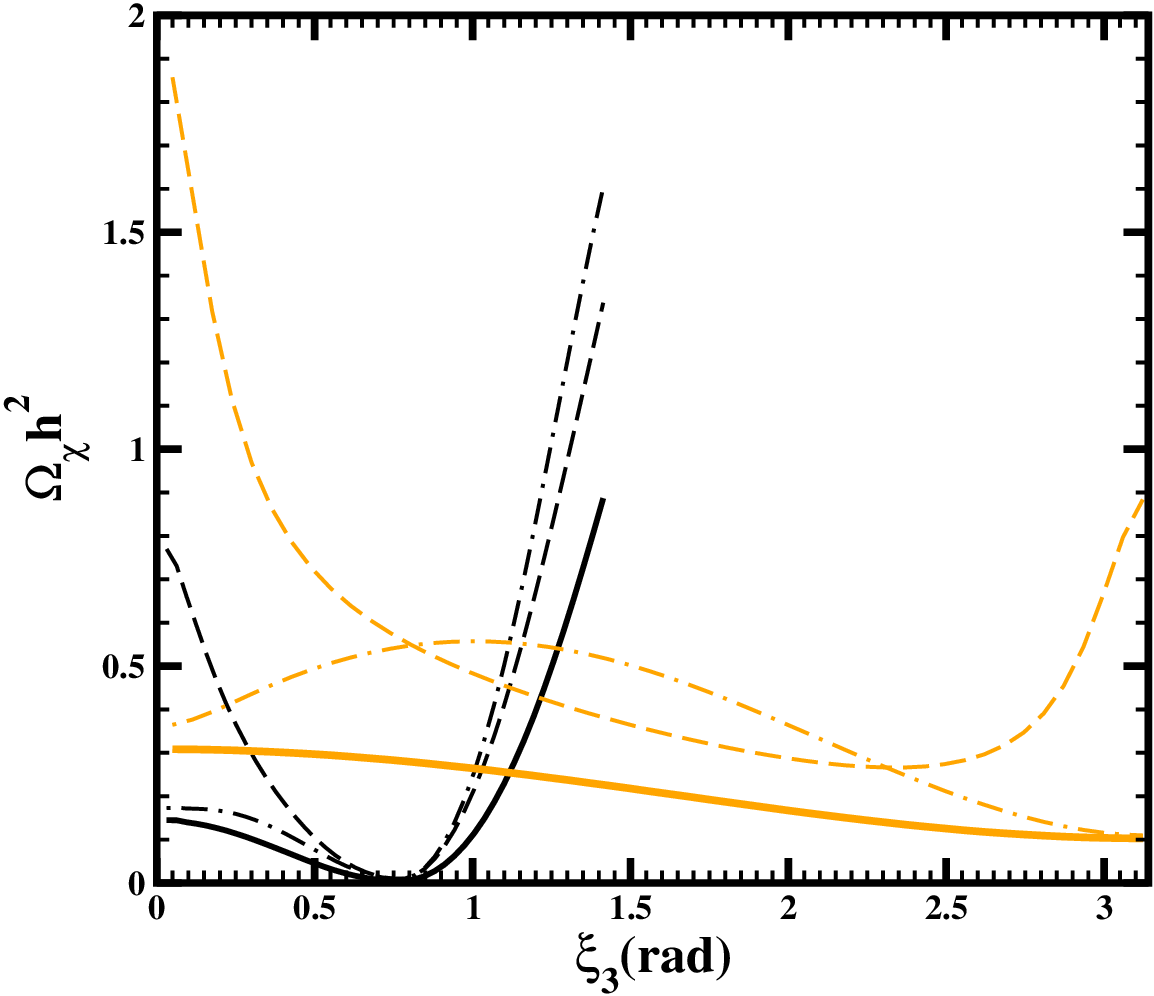}
\includegraphics[width=8cm,height=9cm]{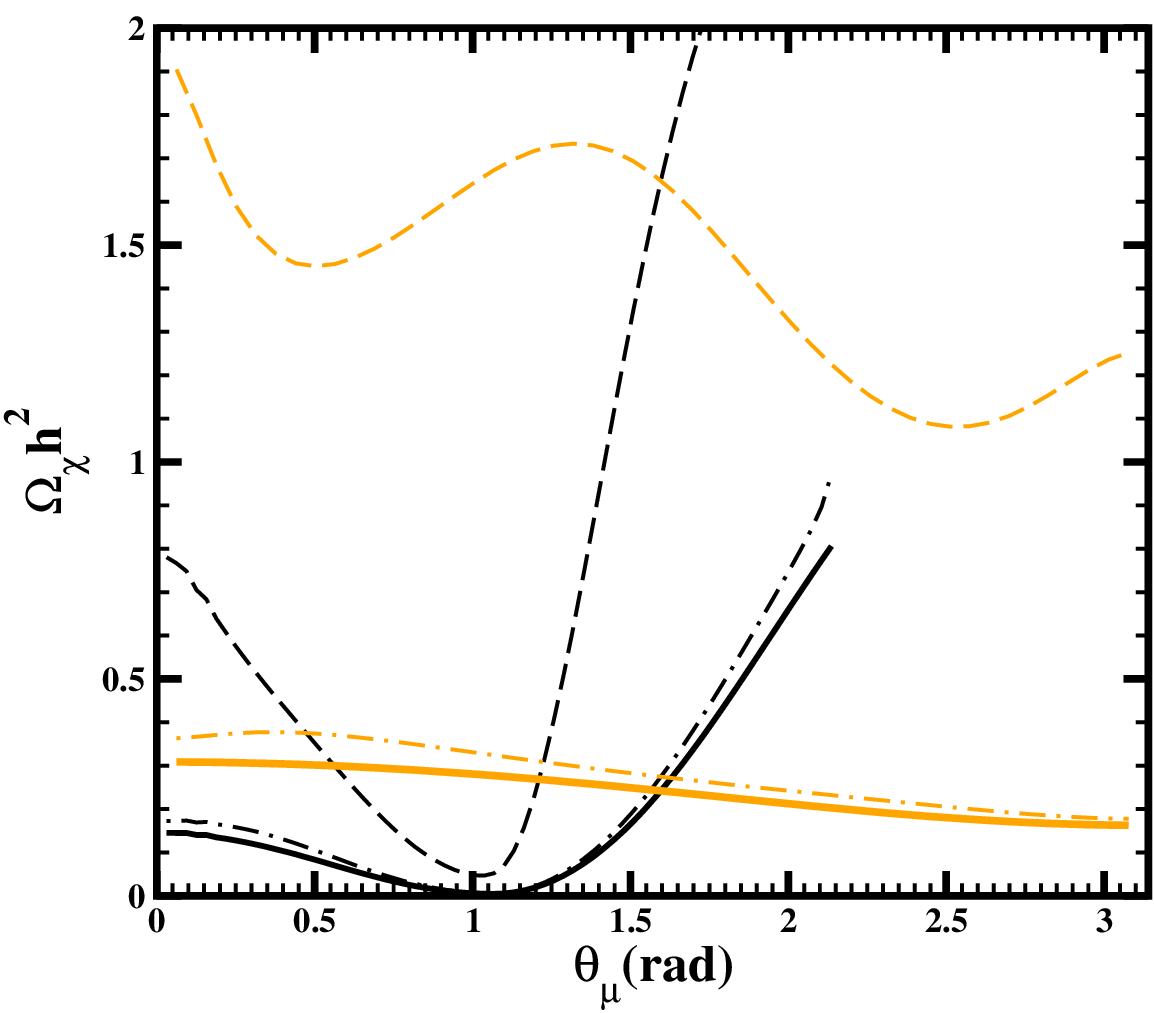}
\caption{A plot of the neutralino relic density $\Omega_{\chi}h^2$ as a function 
of $\xi_3$ and $\theta_{\mu}$ for $m_0=m_{\frac{1}{2}}=A_0=600$ GeV,
$\tan\beta =50$ and using the theoretically predicted value of 
$\Delta m_b$ (black lines), $\Delta m_b=0$ (light lines).  
The contribution of dominant s-channels to the relic density 
are represented by the same type of lines as fig.~(\ref{crtan50}).}
\label{ohtan50}
\end{figure}

\begin{figure}
\hspace*{-0.6in}
\centering
\includegraphics[width=8cm,height=9cm]{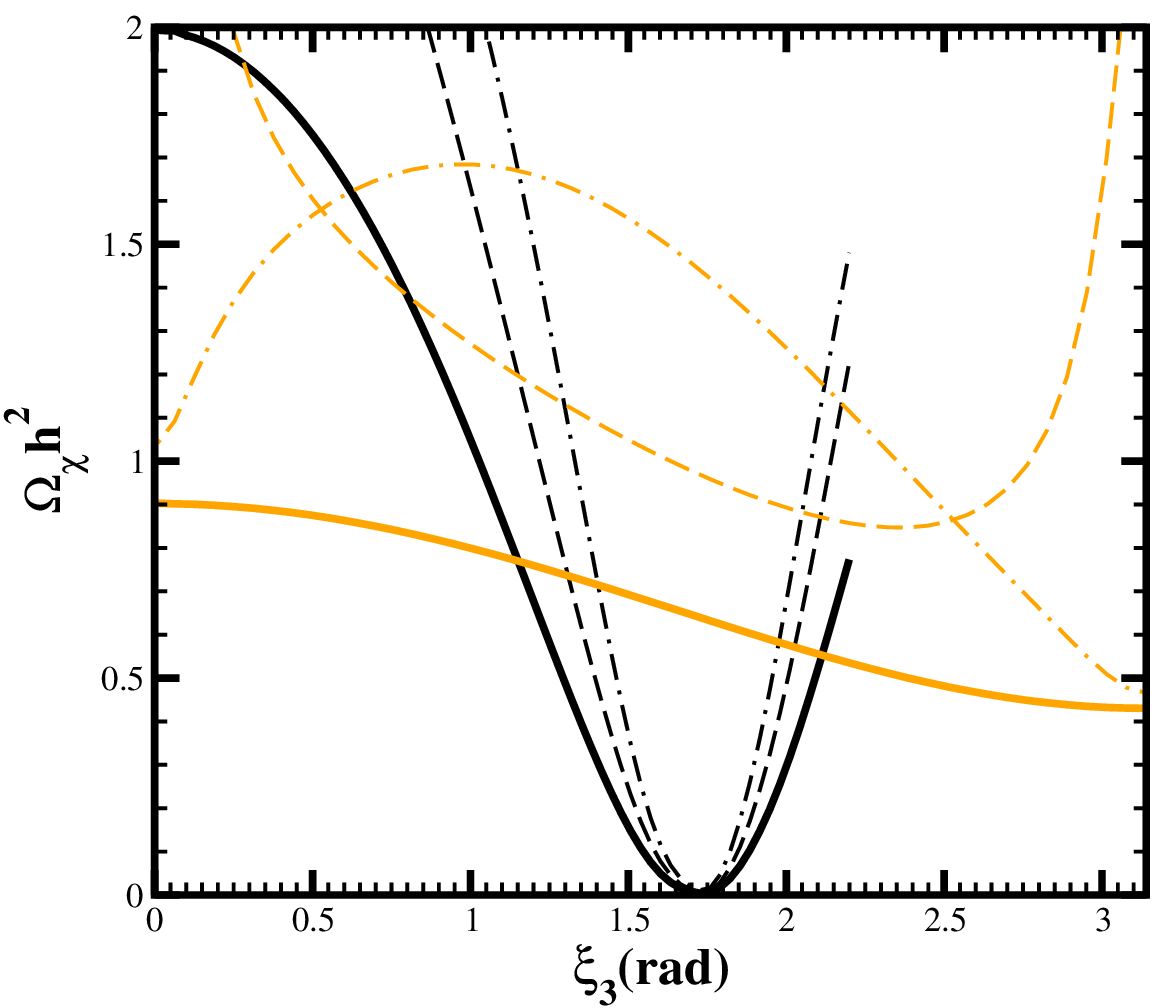}
\includegraphics[width=8cm,height=9cm]{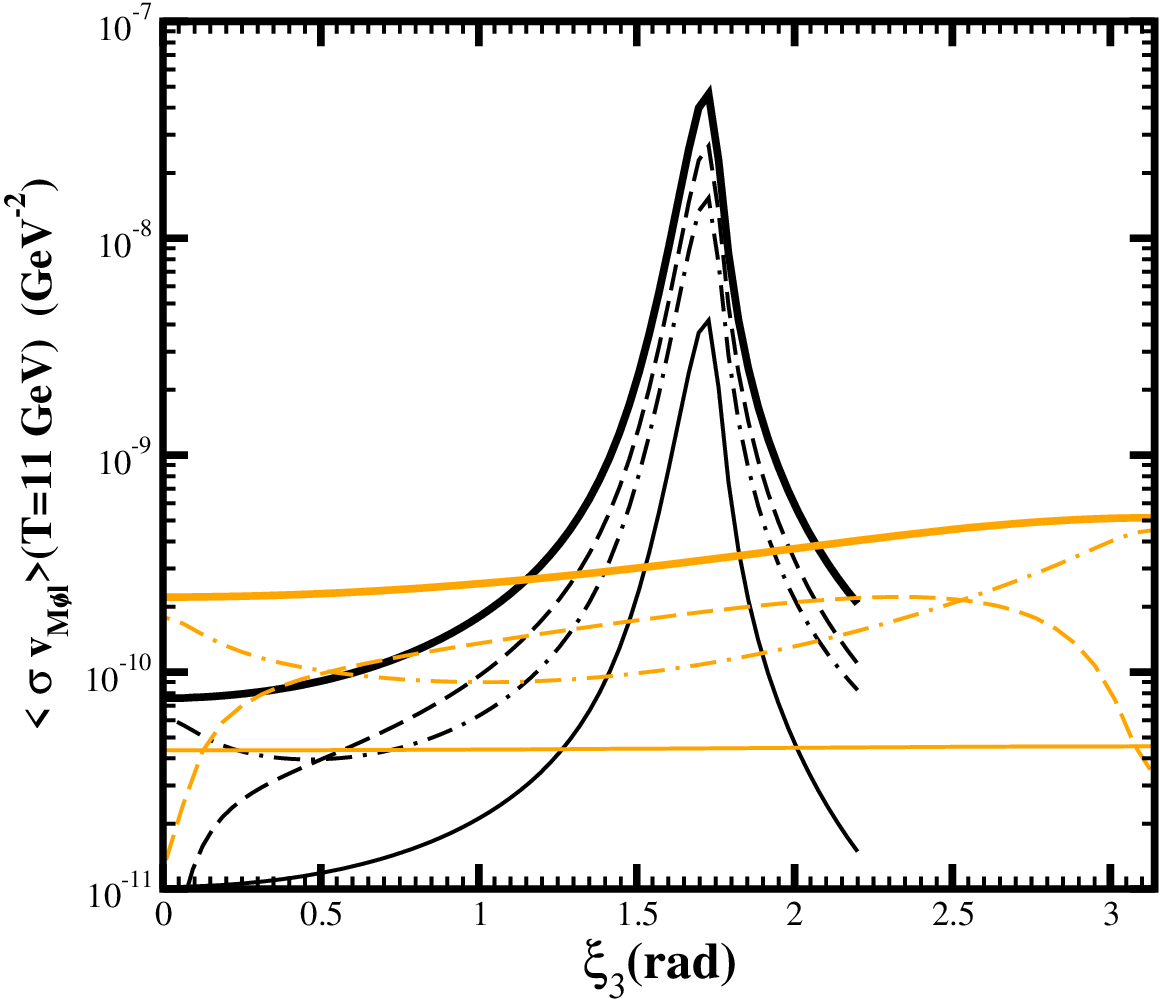}
\caption{Same as Fig.~(\ref{crtan50}) and Fig.~(\ref{ohtan50}) for 
$\tan\beta=40$.}
\label{jan28_graphs.ps_pages2}
\end{figure}

We now give the analysis with inclusion of CP phases. 
In the calculation of the relic density, we only consider the 
contribution from the s-channel exchange of the three Higgs 
$H_1,H_2,H_3$ and the t- and u-channel exchange of sfermions 
as shown in Fig.\ref{ann_diags}. The prediction for the Higgs masses 
and widths 
are extracted from the newly developed software package 
CPsuperH\cite{cpsuperh}. 
The impact of the CP phases on the relic density is as in the 
 case without CP phases, i.e., mainly through $\Delta m_b$. 
On the other hand the effects of the Higgs mixing are marginal. 
In Fig.~(\ref{crtan50}) we give a plot of the thermally averaged 
annihilation cross-section  at a temperature 
of the order of the freeze out temperature $T_f$ 
(see Eq.(\ref{eq:cross}) of the Appendix) as a 
function of $\theta_{\mu}$ and $\xi_3$ for $\tan\beta =50$.  
The contribution of individual channels are also displayed. 
The channels with the $b\bar b$ final state dominate over the 
channels with $\tau\bar\tau$  final state due to the color factor.   
We plot $\Omega_{\chi}h^2$ as a function of $\theta_{\mu}$ and $\xi_3$ 
in Fig.~(\ref{ohtan50}) for the same case. 
Figs. (\ref{crtan50}) and (\ref{ohtan50}) also exhibit the dependence
on the bottom mass correction, as two different values are used; 
the theoretical value of $\Delta m_b$ (black lines) and $\Delta m_b=0$ 
(light lines). 
The large effects of CP phases on the relic density in this case are
clearly evident. In particular it is seen that the largest impact from the 
CP phases arises from their influence on the value of $\Delta m_b$. 
The curves with $\Delta m_b$ 
confined to a constant vanishing value show less variations with the CP 
phases, although $\Omega_{\chi} h^2$ still changes by almost a factor of two due 
to the variation of the Higgs couplings. The large effect of the 
$\Delta m_b$ arises via its effect on $m_A$. 
Similar plots as functions of $\xi_3$  
are given in Fig.~(\ref{jan28_graphs.ps_pages2}) for $\tan\beta =40$.

\begin{table}[t]
\begin{center}
\begin{tabular}{|l|l|l|l|}
\hline
Case & $|d_e| e.cm$ &  $|d_n| e.cm$ &  $C_{Hg} cm$  \\ \hline
(i) & $2.74 \times 10^{-27}$ & $1.79 \times 10^{-26}$ & $8.72 \times 10^{-27}$ 
\\ \hline 
(ii) & $1.29 \times 10^{-27}$ & $1.82 \times 10^{-27}$ & $6.02 \times 10^{-28}$
\\ \hline 
(iii) & $9.72 \times 10^{-28}$
 & $4.19 \times 10^{-26}$ & $1.41 \times 10^{-27}$  
\\ \hline 
\end{tabular}
\end{center}
\caption{The EDMs for $\tan\beta=40$ for cases (i)-(iii) of text.}
\label{edm_value}
\end{table}

\begin{figure}
\centering
\includegraphics[width=7cm,height=7cm]{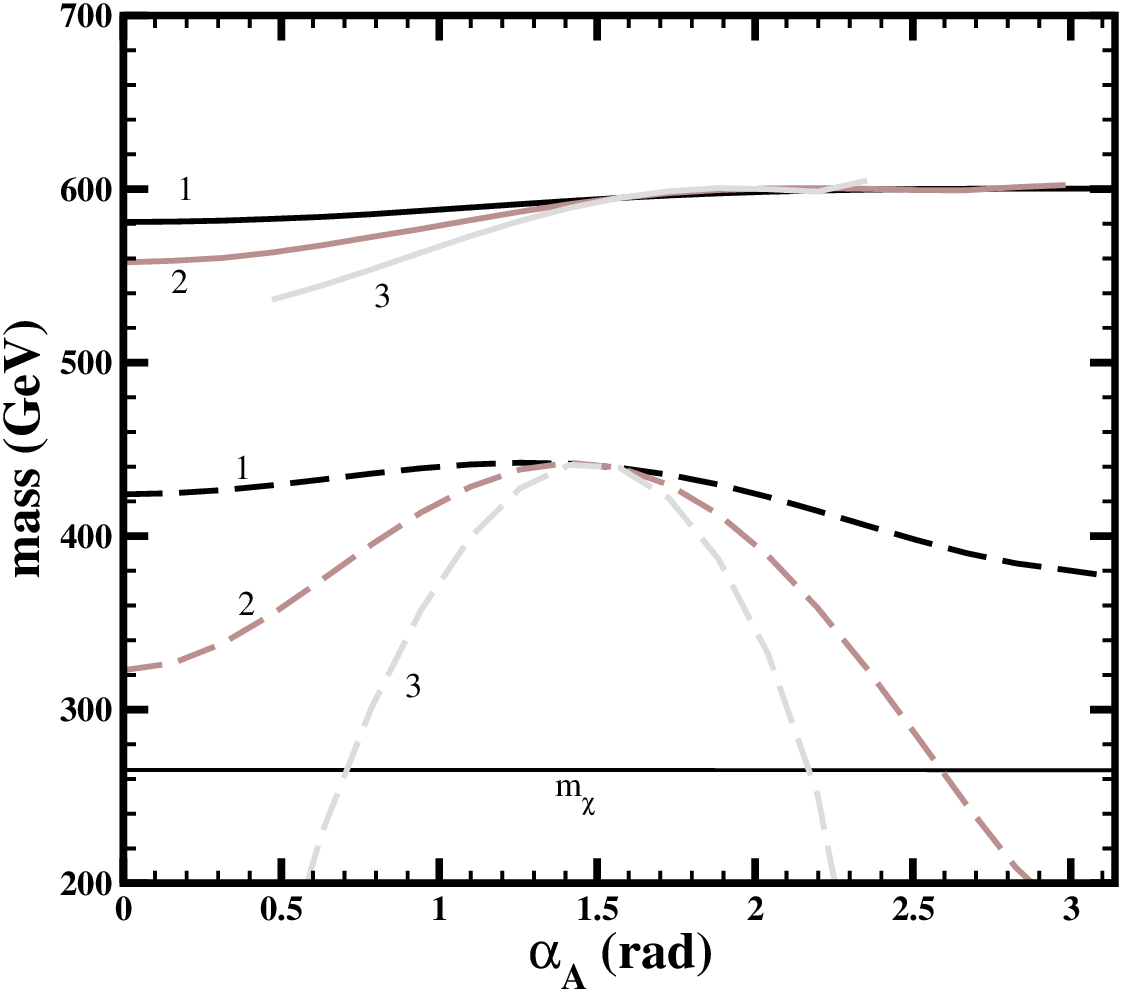}
\includegraphics[width=7cm,height=7cm]{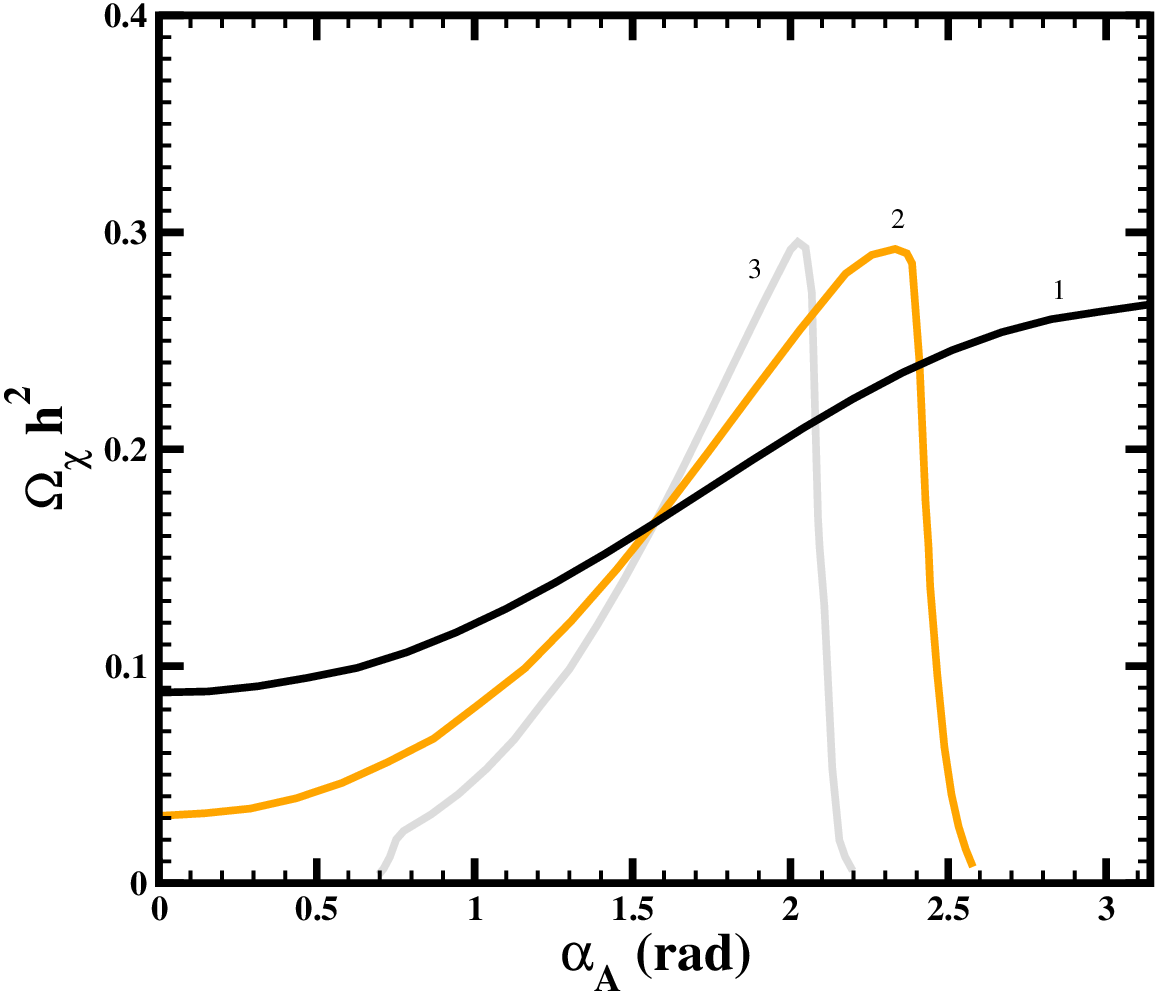}
\caption{ The left graph shows the dependence of 
$m_A$ (solid lines) and $m_{\tilde{\tau_1}}$ (dashed lines) on 
$\alpha_{A_0}$ for $m_0=m_{\frac{1}{2}}=600$~GeV, $\tan\beta=50$ and for 
three different 
values of $|A_0|/m_{\frac{1}{2}}$ (indicated on the curves).
The neutralino relic density for the same three cases is 
displayed  in the graph on the right.}
\label{masspha}
\end{figure}

\begin{figure}[]
\begin{center}
\includegraphics[width=12cm,height=8cm]{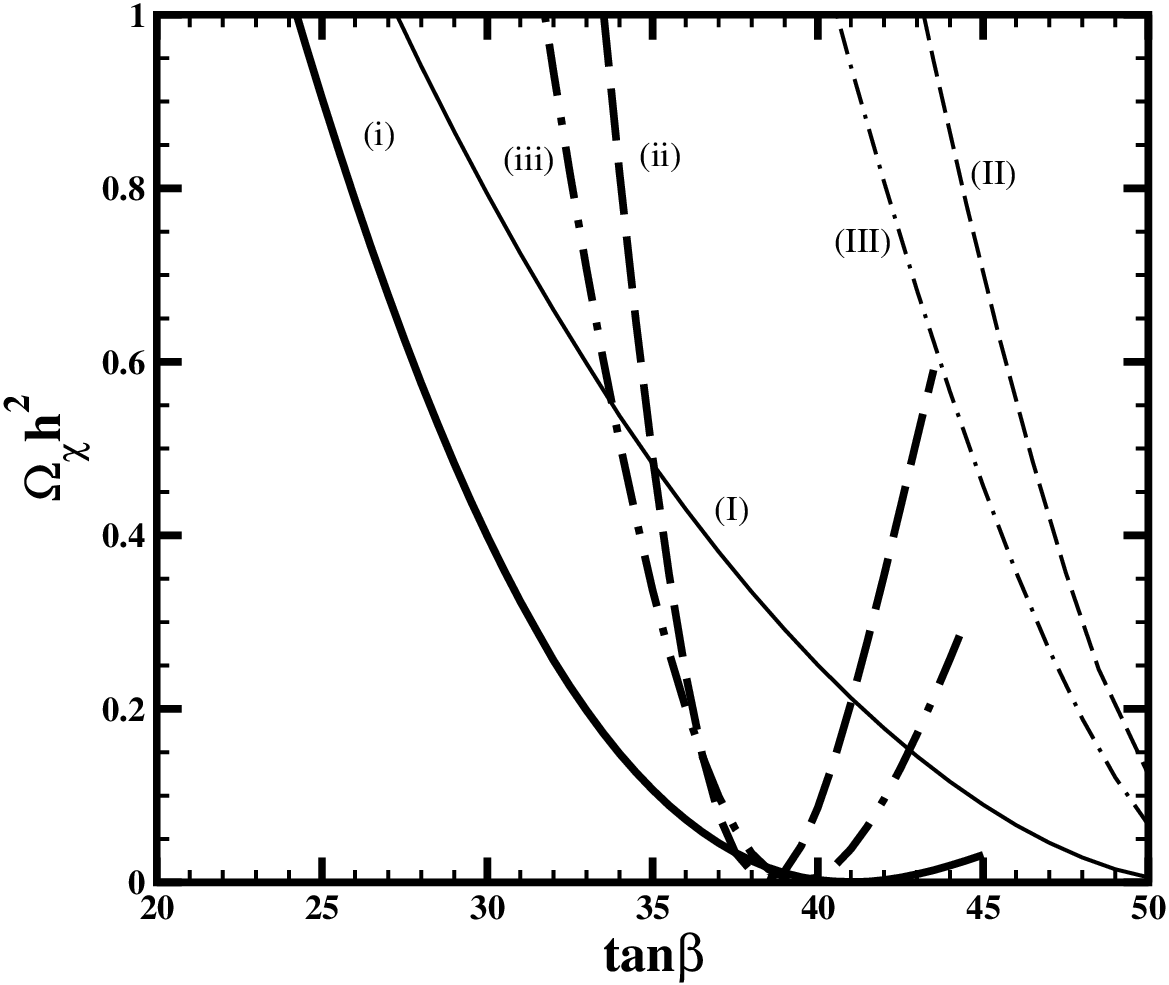}
\end{center}
\caption{The neutralino relic density as a function of $\tan\beta$ for 
the three cases (i), (ii), (iii) of the text. Lines (I), 
(II) and (III) correspond to similar set of SUSY parameters for the case 
of vanishing phases.}   
\label{fig_tanbeta}
\end{figure}

\begin{figure}
\centering
\includegraphics[width=12cm,height=8cm]{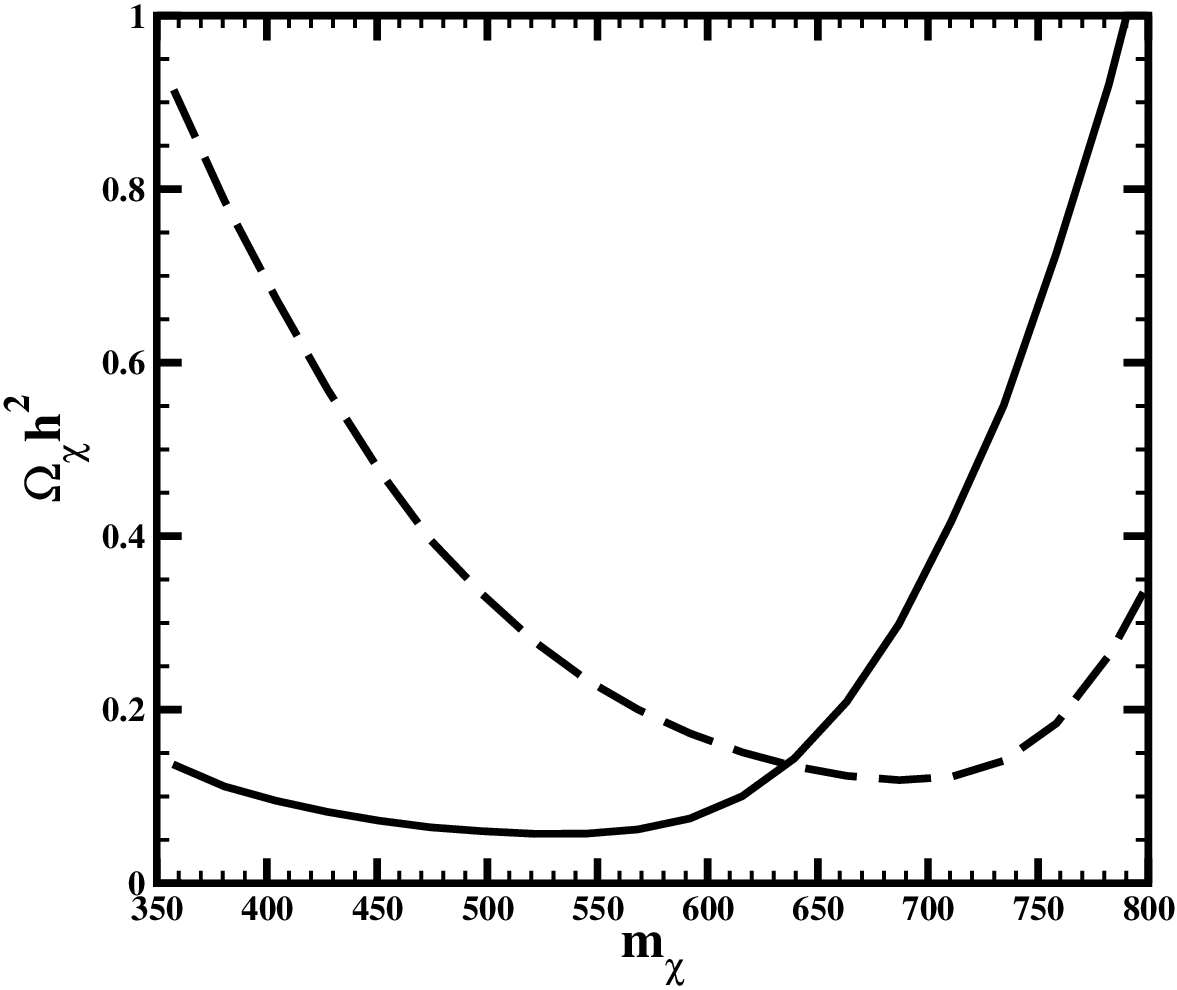}
\caption{The neutralino relic density as a function of the lightest 
neutralino mass for $\tan\beta=30$, $\xi_1=0$, $\xi_2=0$, 
$\xi_3=\pi$, $\alpha_{A_0}=\pi$ and $\theta_\mu=0$. 
Two different values of $m_0$ are displayed:
(i) $m_0=600$ GeV (solid line) and (ii) $m_0=750$ GeV (dashed line).}
\label{fig_lowtb}
\end{figure}

 The dependence of the  neutralino relic density on $\alpha_{A_0}$ 
 is displayed in Fig.~(\ref{masspha}). This dependence arises from
 the effect of $\alpha_{A_0}$ on $m_{\tilde{\tau_1}}$ and $m_A$.
 Thus for fixed $A_0$, variations in $\alpha_{A_0}$ affect 
$m_{\tilde{\tau_1}}$ which can generate $\tilde{\tau}\chi$ 
coannihilations, and even push  $m_{\tilde{\tau}_1}$ below $m_\chi$.
In Fig.~(\ref{fig_tanbeta}) the neutralino relic density is 
displayed as a function 
of $\tan\beta$ for three cases given by: (i) $m_0=m_{1/2}$=$|A_0|=300$ GeV, 
$\alpha_{A_0}=1.0$, 
$\xi_1=0.5$, $\xi_2=0.66$, $\xi_3=0.62$, $\theta_\mu=2.5$;
(ii) $m_0=m_{1/2}=|A_0|=555$ GeV, $\alpha_{A_0}=2.0$, $\xi_1=0.6$, 
$\xi_2=0.65$, $\xi_3=0.65$, $\theta_\mu=2.5$;
\indent (iii) $m_0=m_{1/2}=|A_0|=480$ GeV, $\alpha_{A_0}=0.8$, $\xi_1=0.4$, 
$\xi_2=0.66$, $\xi_3=0.63$, $\theta_\mu=2.5$. 
In all cases the EDM constraints for the electron, the neutron and
for $^{199}Hg$ are satisfied for $\tan\beta=40$ and their values 
are exhibited in table~\ref{edm_value}.
These results may be compared with the current experimental limits on 
the EDM of the electon, the neutron and on  $^{199}Hg$ as follows:
$|d_e|< 4.23 \times 10^{-27}ecm$, $|d_n|< 6.5 \times 10^{-26}ecm$ and 
 $C_{\rm Hg} < 3.0 \times 10^{-26}cm$ from the $^{199}$Hg analysis
(where $C_{\rm Hg}$ is defined as in Ref.~\cite{olive}). 
 From Fig.~(\ref{ohtan}) and Fig.~(\ref{tandel}) 
it is apparent that larger negative corrections to the bottom quark mass 
push the resonance region toward smaller values of $\tan\beta$. 
Use of nonuniversalities for the gaugino masses, including the case 
of having relative signs among them, allows for larger negative 
corrections to the b quark mass. Therefore, it 
is possible to achieve agreement with the WMAP result for lower 
values of $\tan\beta$ than in the mSUGRA case. Considering only 
the main contributions from the gluino-sbottom, the bottom quark 
mass correction will reach its maximum negative value 
for $\theta_\mu + \xi_3 = \pi$. 
The phase of the trilinear coupling also plays a role through the chargino
loop. Thus we investigate the case with $\theta_{\mu}=0$, 
$\xi_3=\pi$, and take $\alpha_{A_0}=\pi$ at the GUT scale.
As an illustration we show in Fig.(\ref{fig_lowtb}) 
that indeed the WMAP result 
is compatible with $\tan\beta=30$ in the resonant s-channel  region. 
The analysis  also implies that the upper limit on the neutralino mass will 
be larger than in the mSUGRA case. For $\tan\beta=30$ we find an upper 
bound of $\sim 700$ GeV, as seen in  Fig.~(\ref{fig_lowtb}).
For comparison the upper bound in mSUGRA is found to be 
500 GeV for $\tan\beta < 30$  in Ref.\cite{wmap1}.

  \section{Conclusion}
   In this paper we have carried out a detailed analysis to study the
   sensitivity of dark matter to the b quark mass. This is done in two
   ways: by assuming that the correction to the b quark mass in 
   a free parameter and also computing it from loop corrections.
   In each case it is found that the relic density is very sensitive 
   to the mass of the b quark for large $\tan\beta$. In the analysis we
   focus on the region where the relic density constraints are
   satisfied by annihilation through resonant Higgs poles. 
   The analysis is then extended to include
   CP phases in the soft parameters taking account of the CP-even and CP-odd
   Higgs mixing. Sensitivity of the relic density to variations
   in the b quark mass and to CP phases are then investigated
    and a great sensitivity to variations in the b quark mass with
    inclusion of phases in again observed. 
   These results have important implications for predictions of 
   dark matter in models where $\tan\beta$ is large, such as in
   unified models  bases on $SO(10)$, and for the observation 
   of supersymmetric dark matter in such models. \\ 
   
\noindent
{\bf Acknowledgments}\\ 
 MEG acknowledges support from the 
`Fundac\~{a}o para a Ci\^encia e Tecnologia' 
under contract SFRH/BPD/5711/2001, the 'Consejer\'{\i}a de Educaci\'on de 
la Junta de Andaluc\'{\i}a' 
and the Spanish DGICYT under contract BFM2003-01266.
The research of TI and PN was supported in part by NSF grant PHY-0139967. 
SS acknowledges support from the European RTN network HPRN-CT-2000-00148. 
\\

\noindent 
{\bf APPENDIX A: Relic Density Analysis}\\
The analysis of neutralino relic density must be done with care since one
has direct channel poles and one must use the accurate method
on doing the thermal averaging over these poles\cite{greist}.
We give here the basic formulas for the relic density 
 analysis\cite{greist,gondolo,Nihei:2002ij}
 \beq 
\Omega_\chi h^2=2.755 \;\ 10^8\times \frac{m_\chi}{{\rm GeV}} Y_0.
\eeq
The evolution equation for $Y$ is given by
\beq
\frac{dY}{dT}=\sqrt{\frac{\pi g_*(T)}{45 G}}
\langle \sigma \vmol \rangle (Y^2-Y_{eq}^2),  
\eeq
Here $\langle \sigma \vmol \rangle$ is the thermal average of the neutralino 
annihilation cross section multiplied by the M{\o}ller velocity 
\cite{gondolo}, $Y_0=Y(T=T_0=2.726 K)$, where $T_0$ is the 
microwave background temperature, $Y_{\rm eq}=Y_{\rm eq}(T)$ is the 
thermal equilibrium abundance given by 
\beq
Y_{\rm eq}(T)=2\times \frac{45}{4\pi^4 h_{\rm eff}(T)}\left(\frac{m_\chi}{T}
\right)^2K_2(\frac{m_\chi}{T}).
\eeq
 The number of degrees of freedom is $g_*\sim 81$. However, we 
use a more precise value as a function of the temperature obtained from 
Ref.~\cite{darksusy} and  the same is done for $h_{\rm eff}$.
To calculate the freeze-out temperature $T_f$ we use the relation
\beq
\frac{d \ln (Y_{\rm eq})}{dT}=\sqrt{\frac{\pi g_*(T)}{45 G}}
\langle \sigma \vmol \rangle Y_{\rm eq}
\delta(\delta+2)
\label{eq:cross}
\eeq
The equation for $Y_0$ is
\beq
\frac{1}{Y_0}=\frac{1}{Y_f}+X_{T_f}
\label{eq:yy}
\eeq
and $Y_f=Y(T_F)=(1+\delta)Y_{\rm eq}(T_f)$. 
We have introduced the amount $X_{T_f}$ such that we can split the 
independent contribution of each channel 
\beq
X_{T_f}=\sqrt{\frac{\pi}{45 G}}\int^{T_f}_{T_0}g_*(T)^{1/2}
\langle \sigma \vmol \rangle dT
\eeq
We have taken $T_0=0$ and $\delta=1.5$ following 
the suggestion of {\it Micromegas}. 
As stated already care must be taken in computing 
thermal averaging since one must integrate over the direct channel
poles properly~\cite{greist}. We use the relation  
\beq
\langle \sigma \vmol \rangle (T)=\frac{1}{8 m_\chi^4 T K_2^2( m_\chi/ T)}
\int^{\infty}_{4 m_\chi^2} ds \sigma(s)(s- 4 m_\chi^2)\sqrt{s} 
K_1\left(\frac{\sqrt{s}}{T}\right)
\eeq
To calculate $\sigma (s)$ from the partial amplitudes we use the following 
definitions
\beq
\sigma(s)=2 w(s)/\sqrt{s(s-4m_\chi^2)}
\eeq
Since we only consider channels $\chi \chi\rightarrow f \bar{f}$, 
($f=b, \tau$), $w(s)$ becomes,
\beq
w(s)=\frac{1}{32 \pi}\sum_{f=b,\tau}c_f\times(1-\frac{4 m_f^2}{s})\tilde{w}_f(s)\eeq
$c_f$ is the color factor so that $c_b=3$, $c_\tau=1$.
The definition of $\tilde{w}_f(s)$ is directly related to the amplitude
\beq
\tilde{w}_f(s)=\frac{1}{2}\int_{-1}^1 dcos\theta_{CM}|A(\chi\chi\rightarrow f
\bar{f})|^2
\eeq


\begin{thebibliography}{999}

\bibitem{bennett}
C.~L.~Bennett {\it et al.},
Astrophys.\ J.\ Suppl.\  {\bf 148}, 1 (2003)
[arXiv:astro-ph/0302207].

\bibitem{spergel}
D.~N.~Spergel {\it et al.},
Astrophys.\ J.\ Suppl.\  {\bf 148}, 175 (2003)
[arXiv:astro-ph/0302209].

\bibitem{dama}
R. Belli et.al., Phys. Lett.{\bf B480},23(2000),
"Search for WIMP annual modulation signature:
results from DAMA/NAI-3 and DAMA/NAI-4 and the global combined
analysis", DAMA collaboration preprint INFN/AE-00/01, 1 February, 2000.

\bibitem{hdms}
L. Baudis, A. Dietz, B. Majorovits, F. Schwamm, H. Strecker and
H.V. Klapdor-Kleingrothaus, Phys. Rev. {\bf D63},022001 (2001).

\bibitem{edelweiss}
A.~Benoit {\it et al.},
Phys.\ Lett.\ B {\bf 545}, 43 (2002)
[arXiv:astro-ph/0206271].

\bibitem{genius}
H.V. Klapdor-Kleingrothaus, et.al., 
"GENIUS, A Supersensitive Germanium Detector System for Rare Events: Proposal", MPI-H-V26-1999,
hep-ph/9910205.

\bibitem{cline}
D.~Cline {\it et al.},
``A Wimp Detector With Two-Phase Xenon,''
Astropart.\ Phys.\  {\bf 12}, 373 (2000).

\bibitem{msugra}
A.H. Chamseddine, R. Arnowitt and P. Nath, \Journal{\PRL}{49}
{970}{1982}; ~R. Barbieri, S. Ferrara and C.A. Savoy, \Journal{\PLB}
{119}{343}{1982}; ~L. Hall, J. Lykken, and S. Weinberg,
\Journal{\PRD}{27}{2359}{1983}:~ P. Nath, R. Arnowitt and A.H. Chamseddine,
\Journal{\NPB}{227}{121}{1983}.
For a recent review see, P.~Nath,
``Twenty years of SUGRA,'' arXiv:hep-ph/0307123.

\bibitem{goldberg}
H.~Goldberg,
Phys.\ Rev.\ Lett.\  {\bf 50}, 1419 (1983);
J.~R.~Ellis, J.~S.~Hagelin, D.~V.~Nanopoulos, K.~A.~Olive and M.~Srednicki,
Nucl.\ Phys.\ B {\bf 238}, 453 (1984).

\bibitem{inmssm}
T.~Ibrahim and P.~Nath,
Phys.\ Rev.\ D {\bf 58}, 111301 (1998).

\bibitem{ccn}
K.L. Chan, U. Chattopadhyay and P. Nath, \Journal {\PRD}{58}{096004}{1998}.
J.~L.~Feng, K.~T.~Matchev and T.~Moroi,
Phys.\ Rev.\ D {\bf 61}, 075005 (2000).

\bibitem{hall}
L.~J.~Hall, R.~Rattazzi and U.~Sarid,
Phys.\ Rev.\ D {\bf 50}, 7048 (1994).
M.~Carena, M.~Olechowski, S.~Pokorski and C.~E.~Wagner,
Nucl.\ Phys.\ B {\bf 426}, 269 (1994).
D.~M.~Pierce, J.~A.~Bagger, K.~T.~Matchev and R.~j.~Zhang,
Nucl.\ Phys.\ B {\bf 491}, 3 (1997).

\bibitem{Ibrahim:2003ca}
T.~Ibrahim and P.~Nath,
Phys.\ Rev.\ D {\bf 67}, 095003 (2003)
[arXiv:hep-ph/0301110].

\bibitem{Ibrahim:2003jm}
T.~Ibrahim and P.~Nath,
Phys.\ Rev.\ D {\bf 68}, 015008 (2003)
[arXiv:hep-ph/0305201].

\bibitem{Ibrahim:2003tq}
T.~Ibrahim and P.~Nath,
arXiv:hep-ph/0311242 (to appear in Phys. Rev. D). 

\bibitem{gomez}
M.E. Gomez, G. Lazarides and C. Pallis, Phys. Rev. {\bf D61}, 
123512 (2000);
Phys.\ Lett.\ B {\bf 487}, 313 (2000);
Nucl.\ Phys.\ B {\bf 638}, 165 (2002)
[arXiv:hep-ph/0203131]; 
Phys.\ Rev.\ D {\bf 67}, 097701 (2003);
C.~Pallis and M.~E.~Gomez,
arXiv:hep-ph/0303098.

\bibitem{Baer:2001yy}
H.~Baer and J.~Ferrandis,
Phys.\ Rev.\ Lett.\  {\bf 87}, 211803 (2001)
[arXiv:hep-ph/0106352];
T.~Blazek, R.~Dermisek and S.~Raby,
Phys.\ Rev.\ Lett.\  {\bf 88}, 111804 (2002)
[arXiv:hep-ph/0107097];
U.~Chattopadhyay, A.~Corsetti and P.~Nath,
Phys.\ Rev.\ D {\bf 66}, 035003 (2002)
[arXiv:hep-ph/0201001];
U.~Chattopadhyay and P.~Nath,
Phys.\ Rev.\ D {\bf 65}, 075009 (2002)
[arXiv:hep-ph/0110341];
S.~Mizuta and M.~Yamaguchi,
Phys.\ Lett.\ B {\bf 298}, 120 (1993)
[arXiv:hep-ph/9208251];
K.~Tobe and J.~D.~Wells,
Nucl.\ Phys.\ B {\bf 663}, 123 (2003)
[arXiv:hep-ph/0301015].

\bibitem{largetan}
M.~E.~Gomez and J.~D.~Vergados,
Phys.\ Lett.\ B {\bf 512}, 252 (2001);
J.~R.~Ellis, T.~Falk, G.~Ganis, K.~A.~Olive and M.~Srednicki,
Phys.\ Lett.\ B {\bf 510}, 236 (2001);
R.~Arnowitt, B.~Dutta and Y.~Santoso,
Nucl.\ Phys.\ B {\bf 606}, 59 (2001)

\bibitem{eedm}
E. Commins, et. al., Phys. Rev. {\bf A50}, 2960(1994).

\bibitem{nedm}
P.G. Harris et.al., Phys. Rev. Lett. {\bf 82}, 904(1999).

\bibitem{atomic}
S.~K.~Lamoreaux, J.~P.~Jacobs, B.~R.~Heckel, F.~J.~Raab and E.~N.~Fortson,
Phys.\ Rev.\ Lett.\  {\bf 57}, 3125 (1986).

\bibitem{deutron}
The edm constraints may be improved further by measurement 
of the deutron edm. See, e.g., 
O.~Lebedev, K.~A.~Olive, M.~Pospelov and A.~Ritz,
arXiv:hep-ph/0402023.

\bibitem{na} 
P. Nath, Phys. Rev. Lett.{\bf 66}, 2565(1991); 
Y. Kizukuri and  N. Oshimo, Phys.Rev.{\bf D46},3025(1992).

 \bibitem{incancel}
T. Ibrahim and P. Nath,  Phys.\ Lett.\ B {\bf 418}, 98 (1998); 
 Phys. Rev. {\bf D57}, 478(1998);
 T. Falk and K Olive, Phys. Lett. {\bf B 439}, 71(1998);
 M. Brhlik, G.J. Good, and G.L. Kane, Phys. Rev. {\bf D59}, 115004
 (1999); A. Bartl, T. Gajdosik, W. Porod, P. Stockinger, and
 H. Stremnitzer,  Phys. Rev. {\bf 60}, 073003(1999);
 S. Pokorski, J. Rosiek and C.A. Savoy, 
 Nucl.Phys. {\bf B570}, 81(2000);
 E.~Accomando, R.~Arnowitt and B.~Dutta,
Phys.\ Rev.\ D {\bf 61}, 115003 (2000);
  U. Chattopadhyay, T. Ibrahim, D.P. Roy, Phys.Rev.D64:013004,2001;
 C.~S.~Huang and W.~Liao,
Phys.\ Rev.\ D {\bf 61}, 116002 (2000);
ibid, Phys.\ Rev.\ D {\bf 62}, 016008 (2000);
 A.Bartl, T. Gajdosik, E.Lunghi, A. Masiero, W. Porod,
H. Stremnitzer and O. Vives, hep-ph/010332;
 M. Brhlik, L. Everett, G. Kane and J. Lykken, Phys. Rev.
 Lett. {\bf 83}, 2124, 1999; Phys. Rev. {\bf D62}, 035005(2000);
  E. Accomando, R. Arnowitt and B. Datta, 
Phys. Rev. {\bf D61},  075010(2000);
T. Ibrahim and P. Nath, Phys. Rev. {\bf D61}, 093004(2000).

\bibitem{olive} 
 T. Falk, K.A. Olive, M. Prospelov, and R. Roiban, Nucl. Phys. 
 {\bf B560}, 3(1999); V.~D.~Barger, T.~Falk, T.~Han, J.~Jiang, T.~Li 
 and T.~Plehn,
Phys.\ Rev.\ D {\bf 64}, 056007 (2001);
S.Abel, S. Khalil, O.Lebedev, Phys. Rev. Lett. {\bf 86}, 5850(2001);
T.~Ibrahim and P.~Nath,
Phys.\ Rev.\ D {\bf 67}, 016005 (2003)

\bibitem{chang}
D. Chang, W-Y.Keung,and A. Pilaftsis, Phys. Rev. Lett. {\bf 82}, 
900(1999). 

\bibitem{bdm2}
K.S. Babu, B. Dutta and R. N. Mohapatra, Phys. Rev. {\bf D61}, 
091701(2000).

\bibitem{pilaftsis}
A. Pilaftsis, Phys. Rev. {\bf D58}, 096010; Phys. Lett.{\bf B435}, 
88(1998);
~A. Pilaftsis and C.E.M. Wagner, Nucl. Phys. {\bf B553}, 3(1999);
~D.A. Demir, Phys. Rev. {\bf D60}, 055006(1999);
~S.~Y.~Choi, M.~Drees and J.~S.~Lee,
Phys.\ Lett.\ B {\bf 481}, 57 (2000);
~M.~Boz,
Mod.\ Phys.\ Lett.\ A {\bf 17}, 215 (2002).

\bibitem{inhiggs1}
T. Ibrahim and P. Nath,  
Phys.Rev.D63:035009,2001; hep-ph/0008237;
T.~Ibrahim,
Phys.\ Rev.\ D {\bf 64}, 035009 (2001);

\bibitem{inhiggs2}
T.~Ibrahim and P.~Nath,
Phys.\ Rev.\ D {\bf 66}, 015005 (2002);
~S.~W.~Ham, S.~K.~Oh, E.~J.~Yoo, C.~M.~Kim and D.~Son,
arXiv:hep-ph/0205244.

\bibitem{Carena:2001fw}
M.~Carena, J.~R.~Ellis, A.~Pilaftsis and C.~E.~Wagner,
Nucl.\ Phys.\ B {\bf 625}, 345 (2002)
[arXiv:hep-ph/0111245].
;
M.~Carena, J.~Ellis, S.~Mrenna, A.~Pilaftsis and C.~E.~Wagner,
arXiv:hep-ph/0211467.

\bibitem{cin}
 U. Chattopadhyay, T. Ibrahim and P. Nath, 
 Phys. Rev. {\bf D60},063505(1999); 
 T. Falk, A. Ferstl and K. Olive, Astropart. Phys. {\bf 13}, 301(2000);
 S. Khalil, Phys. Lett. {\bf B484}, 98(2000);
S. Khalil and Q. Shafi, Nucl.Phys. {\bf B564}, 19(1999);
K. Freese and P. Gondolo, hep-ph/9908390;
 S.Y. Choi, hep-ph/9908397. 

\bibitem{arason}
H. Arason, D.J. Castano, B.E. Kesthelyi, S. Mikaelian,
E.J. Piard, P. Ramond, and B.D. Wright, Phys. Rev. Lett. 
{\bf 67}, 2933(1991);
D. Pierce, J. Bagger, K. Matchev and R. Zhang, Nucl. Phys. {\bf B491},
3(1997)

\bibitem{carena2002}
M.~Carena and H.~E.~Haber,
Prog.\ Part.\ Nucl.\ Phys.\  {\bf 50}, 63 (2003)
[arXiv:hep-ph/0208209].

\bibitem{direct}
K. Greist, Phys. Rev. {\bf D38}, 2357(1988);
J. Ellis and R. Flores, Nucl. Phys. {\bf B307}, 833(1988);
R. Barbieri, M. Frigeni and G. Giudice, Nucl. Phys. {\bf B313}, 725(1989);
A. Bottino et.al., {\bf B295}, 330(1992);
  M. Drees and M.M. Nojiri,
Phys. Rev.{\bf D48},3483(1993); V.A. Bednyakov,
H.V. Klapdor-Kleingrothaus and
S. Kovalenko, Phys. Rev.{\bf D50}, 7128(1994);
P. Nath and R. Arnowitt, Phys. Rev. Lett. {\bf 74}, 4592(1995);
R. Arnowitt and P. Nath, Phys. Rev. {\bf D54}, 2374(1996);
 E. Diehl, G.L. Kane, C. Kolda, J.D. Wells,
 Phys.Rev.{\bf D52}, 4223(1995);
 L. Bergstrom and P. Gondolo, Astrop. Phys. {\bf 6}, 263(1996);
 H. Baer and M. Brhlik, Phys.Rev.{\bf D57},567(1998);
J.D. Vergados, Phys. Rev. {\bf D83}, 3597(1998);
J.L. Feng, K. T. Matchev, F. Wilczek, Phys.Lett. {\bf B482}, 388(2000);
M.~Brhlik, D.~J.~Chung and G.~L.~Kane,
Int.\ J.\ Mod.\ Phys.\ D {\bf 10}, 367 (2001);  V.A. Bednyakov and
H.V. Klapdor-Kleingrothaus,  Phys.\ Rev.\ D {\bf 63}, 095005 (2001);
M.~E.~Gomez and J.~D.~Vergados, Phys.\ Lett.\ B {\bf 512}, 252 (2001);
A. Corsetti  and P. Nath, 
Phys.\ Rev.\ D {\bf 64}, 125010 (2001);
A.~B.~Lahanas, D.~V.~Nanopoulos and V.~C.~Spanos,
Phys.\ Lett.\ B {\bf 518}, 94 (2001);
J.L. Feng, K. T. Matchev, F. Wilczek, Phys.\ Rev.\ D {\bf 63}, 045024 (2001);
V.~D.~Barger and C.~Kao, Phys.\ Lett.\ B {\bf 518}, 117 (2001).

\bibitem{gomez1}
D.~G.~Cerdeno, E.~Gabrielli, M.~E.~Gomez and C.~Munoz,
JHEP {\bf 0306}, 030 (2003)
[arXiv:hep-ph/0304115].

\bibitem{Ellis:2003si}
J.~R.~Ellis, K.~A.~Olive, Y.~Santoso and V.~C.~Spanos,
arXiv:hep-ph/0310356.

\bibitem{micromegas}
G.~Belanger, F.~Boudjema, A.~Pukhov and A.~Semenov,
Comput.\ Phys.\ Commun.\  {\bf 149}, 103 (2002)
[arXiv:hep-ph/0112278].
For an analysis of uncertainties in the determination of relic density 
in  different public codes see, 
B.~C. Allanach,  G.~Belanger, F.~Boudjema, A.~Pukhov and W.~Porod,
hep-ph/0402161.

\bibitem{wmap1}
J.~R.~Ellis, K.~A.~Olive, Y.~Santoso and V.~C.~Spanos,
Phys.\ Lett.\ B {\bf 565}, 176 (2003)
[arXiv:hep-ph/0303043].

\bibitem{wmap2}
H.~Baer and C.~Balazs,
JCAP {\bf 0305}, 006 (2003)
[arXiv:hep-ph/0303114].;
U.~Chattopadhyay, A.~Corsetti and P.~Nath,
Phys.\ Rev.\ D {\bf 68}, 035005 (2003)
[arXiv:hep-ph/0303201];
H.~Baer, C.~Balazs, A.~Belyaev, T.~Krupovnickas and X.~Tata,
JHEP {\bf 0306}, 054 (2003)
[arXiv:hep-ph/0304303];
A.~B.~Lahanas and D.~V.~Nanopoulos,
Phys.\ Lett.\ B {\bf 568}, 55 (2003)
[arXiv:hep-ph/0303130];
H.~Baer, T.~Krupovnickas and X.~Tata,
JHEP {\bf 0307}, 020 (2003)
[arXiv:hep-ph/0305325];
A.~B.~Lahanas, N.~E.~Mavromatos and D.~V.~Nanopoulos,
Int.\ J.\ Mod.\ Phys.\ D {\bf 12}, 1529 (2003)
[arXiv:hep-ph/0308251].

\bibitem{baerhiggspole}
H.~Baer and J.~O'Farrill,
arXiv:hep-ph/0312350.

\bibitem{cpsuperh} 
J.S. Lee, A. Pilaftsis, M. Carena, S.Y. Choi, M. Drees, J. Ellis, 
C.E.M. Wagner, Comput.Phys.Commun. 156 (2004) 283-317.

\bibitem{greist}
K. Greist and D. Seckel, Phys. Rev. {\bf D43}, 3191(1991);
R. Arnowitt and P. Nath, Phys. Lett. {\bf B299}, 103(1993); 
 Phys. Rev. Lett. {\bf 70}, 3696(1993);
 H. Baer and M. Brhlik, Phys. Rev. {\bf D53}, 597(1996); 
 V. Barger and C. Kao, Phys. Rev. {\bf D57}, 3131(1998).

\bibitem{gondolo}
P. Gondolo and G. Gelmini, Nucl. Phys. {\bf B360}, 145(1991).

\bibitem{Nihei:2002ij}
T.~Nihei, L.~Roszkowski and R.~Ruiz de Austri,
JHEP {\bf 0203}, 031 (2002)
[arXiv:hep-ph/0202009].

\bibitem{darksusy}
P.~Gondolo, J.~Edsjo, P.~Ullio, L.~Bergstrom, M.~Schelke and E.~A.~Baltz,
arXiv:astro-ph/0211238. http://www.physto.se/~edsjo/darksusy/

\end{thebibliography}
\end{document}